\documentclass{article}
\usepackage{placeins}
\usepackage{float}
\usepackage{arxiv}
\usepackage{amsmath}
\usepackage[utf8]{inputenc} 
\usepackage[T1]{fontenc}    
\usepackage{hyperref}       
\usepackage{url}            
\usepackage{booktabs}       
\usepackage{amsfonts}       
\usepackage{nicefrac}       
\usepackage{microtype}      
\usepackage{graphicx}
\usepackage{caption}
\usepackage{array}
\usepackage{subfig}
\usepackage{tabularx}
\usepackage{multirow}
\usepackage{csquotes}  
\usepackage{lscape}
\usepackage{color}
\usepackage{xcolor}
\usepackage[linesnumbered,ruled,vlined]{algorithm2e}
\renewcommand{\arraystretch}{1.25}
\usepackage{amsmath,amsfonts,tikz-cd}
\usepackage{tikz}
\usepackage{pgfplots}
\usetikzlibrary{decorations.pathreplacing,angles,quotes}

\title{The $MAP/M/s+G$ Call Center Model with General Patience Times: Stationary Solutions and First Passage Times}

\author{
	 Omer~Gursoy \\
	Electrical and Electronics Engineering Dept.\\
	Bilkent University\\
	Ankara, Turkey \\
	\texttt{gursoy@ee.bilkent.edu.tr} \\
	\And
  Kamal A.~Mehr
 \thanks{This research was partially carried out when K. A.~Mehr paid a one-year visit to Bilkent University in 2017 as a Ph. D. student.} 
\\
  University of Tabriz\\
  \texttt{ka.ad.me@gmail.com} \\
   \And
  Nail~Akar \\
  Electrical and Electronics Engineering Dept.\\
  Bilkent University\\
  Ankara, Turkey \\
  \texttt{akar@ee.bilkent.edu.tr} \\
}

\begin{document}
\maketitle

\begin{abstract}
We study the $MAP/M/s+G$ queuing model with MAP (Markovian Arrival Process) arrivals, exponentially distributed service times, infinite waiting room, and generally distributed patience times. Using sample-path arguments, we propose to obtain the steady-state distribution of the virtual waiting time and subsequently the other relevant performance metrics of interest for the $MAP/M/s+G$ queue by means of finding the steady-state solution of a properly constructed Continuous Feedback Fluid Queue (CFFQ). The proposed method is exact when the patience time is a discrete random variable and is asymptotically exact when it is continuous/hybrid for which case discretization of the patience time distribution and subsequently the steady-state solution of a Multi-Regime Markov Fluid Queue (MRMFQ) is required. 
Besides the steady-state distribution, we also propose a new method to approximately obtain the first passage time distribution for the virtual and actual waiting times in the $MAP/M/s+G$ queue. Again, using sample-path arguments, finding the desired distribution is also shown to reduce to obtaining the steady-state solution of a larger dimensionality CFFQ where the deterministic time horizon is to be approximated by Erlang or Concentrated Matrix Exponential (CME) distributions. Numerical results are presented to validate the effectiveness of the proposed method. 

\end{abstract}


\section{Introduction}
\label{introduction}
In call centers, customer impatience refers to a situation when a customer who gets delayed in the queue runs out of patience and subsequently abandons the system when the queue wait exceeds the so-called patience (or abandonment) time. Impatience is a well-known phenomenon in call centers and a number of queuing models have been proposed in the literature to model call centers with impatience; see for example Aksin et al.  \cite{aksin_etal_POM07}, 
Brown at al. \cite{brown_etal_JASA05},
Gans, Koole, and Mandelbaum \cite{Gans.2003}, 
Jouini, Koole, and Roubos \cite{jouini_etal_IIE13} and the references therein for a survey of call centers along with a collection of such queuing models. The variations in these models stem from the assumptions on customer interarrival and service times, number of waiting rooms, and more specifically, the abandonment time distribution. 

The most basic queuing model with impatience is the $M/M/s+M$ model (also known as the Palm model \cite{palm_tele53} or the Erlang-A model) with Poisson customer arrivals, exponentially distributed service times, infinite waiting room, and each customer is assigned an exponentially distributed abandonment time. The reader is referred to Garnett, Mandelbaum, and Reiman \cite{Garnett.2002} for exact and approximate expressions for certain performance measures of the Erlang-A model.
The $M/M/s/k+M$ queue with $k$ waiting rooms where the number of customers in the system over time is a birth-and-death process, is investigated by Whitt \cite{Whitt.1999}.
However, existing statistical call center traffic analysis reveals that abandonment times are of non-exponential nature Brown at al. \cite{brown_etal_JASA05}, Jouini at al. \cite{jouini_etal_IIE13}. Moreover, higher order moments of the patience time are known to play a major role in the overall performance of call centers; see Whitt \cite{Whitt.2005}.
In the work \cite{Baccelli.1981b} by Baccelli and Hebuterne, arriving customers know about their queue wait
at their arrival epochs and they abandon immediately at their arrival epoch if the queue wait exceeds a generally distributed patience time. This model coincides with the classical $M/M/s+G$ model for those performance measures of interest not relying on queue lengths such as the abandonment probability, queue wait distributions, etc. The two works by Brandt and Brandt \cite{Brandt.1999},\cite{Brandt.2002} study the more general $M(k)/M(k)/s+G$ system where the arrival
and service rates are allowed to depend on the number of customers in the system. The authors assume that the customers abandon at the end of their patience and therefore queue-length related performance measures can also be obtained. 
The references Zelytn \cite{zelytn} and Zalytn and Mandelbaum \cite{Zeltyn.2005} present the existing exact results on the $M/M/s+G$ queue along with approximate many-server asymptotic results whereas the reference by Mandelbaum and Zelytn \cite{Mandelbaum.2004} presents empirically-driven experiments for the same model. Whitt \cite{Whitt.2005} develops an engineering approach to provide approximate results for the $M/G/s/k+G$ for several performance metrics of interest. On the other hand, Movaghar \cite{Movaghar.1998} presents a generalized solution for the offered waiting time, the probability of missing deadline, and the probability of blocking for the $M(n)/M(n)/s/FCFS+G$ queue for both infinite and finite queue capacity cases. 

Markovian Arrival Process (MAP) refers to a very general class of Markov modulated processes used for modeling interarrival times in queueing systems; see Lucantoni, Meier-Hellstern, and Neuts \cite{lucantoni_adv90}, Lucantoni \cite{lucantoni_sm91},
Neuts \cite{neuts89},
Ramaswami and Latouche \cite{ramaswami.latouche.02}.
It is shown by Asmussen and Koole \cite{asmussen_koole_jap93} that the set of MAPs is dense in the set of all stationary point processes which makes the MAP a very powerful tool for modeling interarrival times in queueing systems. In this regard, MAPs have been successfully used for fitting real workload traces with short-range or long-range dependent behavior Andersen and Nielsen \cite{andersen_nielsen_jsac98},
Casale \cite{casale_sigmetrics},
Casale, Zhang, and Smirni \cite{casale_etal_peva10},
Okamura, Dohi, and Trivedi \cite{okamura_etal_ton09}.
MAPs have also been used to model queueing systems arising specifically in call centers. 
Asmussen and M{\o}ller \cite{asmussen_moller_questa01} study the stationary waiting time distribution for the $MAP/M/s$ queue without abandonments. The $MAP/M/s+D$ queue with deterministic abandonments is thoroughly investigated by Choi, Kim, and Zhou \cite{Choi.2004} who first derive the virtual waiting time distribution and subsequently the loss probability and the actual waiting time distribution. The virtual waiting time results of Choi et al. \cite{Choi.2004} are refined by Kawanishi and Takine \cite{Kawanishi.2016} in which the stationary queue length distribution is also obtained for the $MAP/M/s+D$ queue. A multi-server queueing system with finite buffer and impatient heterogeneous MAP customer arrivals with exponentially distributed patience times is investigated Kim et al. \cite{kim_etal_AMM13}.


In this paper, we consider the $MAP/M/s+G$ queue with infinite waiting room whose steady-state solution is an open problem to the best of our knowledge for generally distributed patience times. Using sample-path arguments, obtaining the steady-state distribution of the virtual waiting time is shown in this paper to reduce to finding the steady-state solution of a Continuous Feedback Fluid Queue (CFFQ). Subsequently, other performance measures of interest not relying on queue lengths such as the abandonment probability, the zero wait probability, actual waiting time distribution of successful calls, etc. can be derived from the virtual waiting time distribution.
The proposed method is exact when the abandonment time is a discrete random variable and is asymptotically exact when the patience time is a continuous or hybrid random variable in which case discretization of the abandonment time distribution is required. With such discretization, CFFQs reduce to the so-called Multi-Regime Markov Fluid Queues (MRMFQ) for which numerically efficient and stable steady-state solvers are available Kankaya and Akar \cite{kankaya.2008}, da Silva Soares and Latouche \cite{soares_latouche}. The computational complexity of the proposed method depends linearly on the number of discretization levels and results of arbitrary accuracy is within reach when the patience time is a continuous random variable. 
One of the main contributions of this paper is that we extend the MRMFQ-based approach taken by Van Houdt \cite{vanhoudt_ejor12} for a single-server queue to a multi-server $MAP/M/s+G$ call center system in case of general patience time distributions. As the numerical engine, we use the steady-state MRMFQ-solver proposed by Kankaya and Akar \cite{kankaya.2008} that relies on the ordered Schur decomposition.

Most existing studies consider the steady-state solution of call center models. However, obtaining transient performance measures of interest is also key for call center models Knessl and Van Leeuwaarden \cite{knessl_mmor15}. 
First passage times are known to be among major transient performance measures and we believe that they can potentially be used for risk analysis in dynamic provisioning of call centers. 
In this paper, we first study the first passage time distribution for the virtual waiting time for the $MAP/M/s+G$ queue. Then, by extending the method that we developed for the virtual waiting time, a queueing model is also obtained for the first passage time distribution for the actual waiting time.
There have been several studies in the recent literature to approximately obtain transient quantities of interest in certain queueing systems via the steady-state solution of properly constructed stochastic systems using sample path arguments. The reference Van Houdt and Blondia \cite{van2005}
obtains the transient queue lengths and waiting time distributions using steady-state analysis for a discrete-time queue. Velthoven, Van Houdt, and Blondia \cite{velthoven_etal_qest07} provide an algorithm to find certain transient performance measures for every possible initial configuration of a Quasi-Birth-and-Death (QBD) Markov chain by means of the steady-state solution of another properly constructed Markov chain. 
Similarly, a numerical method has been proposed by Yazici and Akar \cite{yazici_akar_anor17} for finding the ruin probabilities for a general continuous-time risk problem using the steady-state solution of a certain MFQ using Erlangization for keeping track of the deterministic time horizon where an Erlang distribution of order $\ell$ is used to approximate a deterministic quantity.
A recent work by Akar et al. \cite{akar_mcap_submitted} obtains transient and first passage time distributions of first- and second-order MRMFQs using Concentrated Matrix Exponential (CME) distributions as opposed to Erlangization for approximating deterministic time horizons.
Motivated by this line of research, as the second main contribution of this paper, the first passage time distribution for both the virtual and actual waiting times for the $MAP/M/s+G$ call center model is shown to be computable through the steady-state analysis of an auxiliary MRMFQ of larger order, using both Erlangization and CME distributions.

The rest of the paper is organized as follows. In Section~\ref{MRMFQ}, preliminaries on CFFQs and MRMFQs and their steady-state solutions are presented. The $MAP/M/s+G$ queue and its steady-state solution using the theory of MRMFQs is presented in Section~\ref{stationary}. The approximate solution for the first passage time distribution for the virtual and actual waiting times in the $MAP/M/s+G$ queue are given in Section~\ref{firstpassagesection}. Validation of the proposed algorithms is done with a number of numerical examples along with comparisons with simulations in Section~\ref{numerical}. Finally, we conclude.

\section{Multi-Regime Markov Fluid Queues}
\label{MRMFQ}
Conventional Markov Fluid Queues (MFQ) are described by a joint Markovian process ${\mathbf X(t)}= (X_f(t),X_m(t))$, $t\geq 0,$ where $0\leq X_f(t) \leq B$ represents the fluid level in the buffer, $B$ denotes the buffer capacity, and the modulating phase process $X_m(t) \in \{1,2,\ldots,n\}$ is an $n$-dimensional Continuous Time Markov Chain (CTMC) with generator $Q$. 
In MFQs, the net rate of fluid change (or drift) is $r_i$ when the phase of the modulating process $X_m(t)$ is $i$. The drift matrix $R$ is the diagonal matrix of drifts: $ R=\mathbf{diag}\{
r_1, r_2, \ldots, r_{n} \}$ and the process ${\mathbf X(t)}$ is fully characterized with the pair $(Q,R)$.
In the seminal work of Anick, Mitra, and Sondhi \cite{anick_mitra82}, the stationary solutions of MFQs are studied for the infinite buffer capacity case using a spectral expansion approach whereas Tucker \cite{tucker88} extends this analysis to the finite buffer capacity case. A more recent work by Akar and Sohraby \cite{akar_sohraby_jap04} obtains the stationary solution of MFQs for both infinite and finite buffers using an alternative additive decomposition technique which is computationally more stable and efficient than spectral decomposition. Transient solution of MFQs are studied iby Ahn and Ramaswami \cite{ahn_ram_JAP2005} and Ahn, Badescu, and Ramaswami \cite{ahn_etal_questa07} using matrix-analytic techniques.

Recently, there has been a surge of interest in MFQs for which the drift not only depends on $X_m(t)$ but also on the instantaneous fluid level $X_f(t)=x$ and moreover
the generator for the phase process $X_m(t)$ also depends on the latter. 
In Continuous Feedback Fluid Queues (CFFQ), this dependence is continuous and CFFQs are fully characterized with a pair of level dependent generator and drift matrices $(Q(x),R(x))$; see Scheinhadrt, Van Foreest, and Mandjes \cite{cffq}. However, exact stationary solutions for CFFQs are only available for special cases and approximative techniques are required for their solutions. Another class of MFQs is Multi-Regime MFQs (MRMFQ) for which the buffer is partitioned into a finite number of non-overlapping regimes (intervals) and the drift depends on the phase as well as the regime of the fluid level and also the generator for the phase process $X_m(t)$ depends on the regime; see Mandjes, Mitra, and Scheinhardt \cite{mandjes_mitra} and Kankaya and Akar \cite{kankaya.2008} for a more elaborate discussion of MRMFQs and their stationary solutions. In particular, the buffer is partitioned into $K > 1$ regimes with the boundaries $0=T^{(0)}<T^{(1)}< \cdots <T^{(K-1)}<T^{(K)}=B$. 
When $T^{(k-1)}<X(t)<T^{(k)}$, the fluid process is said to be in regime $k$ at time $t$. 
The MRMFQ is subsequently characterized with a level-dependent pair of matrices $(Q(x),R(x))$ which turn out to have the following form 
\begin{equation}
Q(x) = 
\begin{cases} 
Q^{(k)} & \text{if } T^{(k-1)}<x<T^{(k)}, \quad k=1,2,\ldots,K, \\
\tilde{Q}^{(k)} & \text{if } x = T^{(k)} ,\quad  k=0,1,\ldots,K,\\
\end{cases}
\label{generatorx}
\end{equation} 
\begin{equation}
R(x) = 
\begin{cases} 
R^{(k)} & \text{if } T^{(k-1)}<x<T^{(k)}, \quad k=1,2,\ldots,K, \\
\tilde{R}^{(k)} & \text{if } x = T^{(k)} , \quad 0,1,\ldots, K,\\
\end{cases}
\label{driftx}
\end{equation}
where the regime-$k$ generator and drift matrices are denoted by $Q^{(k)}$ and $R^{(k)}$, respectively, and the boundary-$k$ generator and drift matrices are 
denoted by $\tilde{Q}^{(k)}$ and $\tilde{R}^{(k)}$, respectively. 
It is clear that strictly negative (strictly positive) fluid drifts at boundary 0 (boundary $K$) are not allowed for the purpose of restricting the fluid level to the interval $[0,B]$.
The regime-$k$ joint probability density function (pdf) vector $f^{(k)}(x)$ is defined as follows:
\begin{equation}
f^{(k)}(x) =  \{ f_1^{(k)}(x), f_2^{(k)}(x),\ldots,f_{n}^{(k)}(x) \} \,   \label{density}
\end{equation}
where
\begin{equation}
f_i^{(k)}(x)  =  \lim\limits_{t\to \infty } \frac{d}{dx}  \Pr\{X_f(t)\leq x,\: X_m(t)=i\}, \quad T^{(k-1)}<x<T^{(k)}. 
\end{equation}
Similarly, the steady-state boundary-$k$ probability mass accumulation (pma) vector $c^{(k)}$ is defined as follows:
\begin{equation}
c^{(k)}  = \{ c_1^{(k)}, c_2^{(k)}, \ldots, c_{n}^{(k)}  \} ,  \quad
c_i^{(k)}  =  \lim\limits_{t\to \infty } \Pr\{X_f(t)=T^{(k)},\; X_m(t)=i\}, \quad 0\leq k\leq K .\label{accumulation}
\end{equation}
A matrix-analytic algorithm has been proposed by Kankaya and Akar \cite{kankaya.2008} to obtain the
joint pdf vector given in (\ref{density}) in matrix exponential form and the joint pma vector in (\ref{accumulation}).
This numerical algorithm requires the solution of a linear matrix equation of at most size $n(2K+1)$ for an MRMFQ with $n$ phases and $K$ regimes. The computational complexity of the proposed algorithm can be reduced to ${\cal O}(n^3 K)$ on the basis of the observation that the linear matrix equation is in block tridiagonal form; see Yazici and Akar \cite{yazici_akar_peva13}. Moreover, it is shown in this work that MRMFQ-based models can be used to approximately obtain the stationary solution for more general CFFQs through the discretization of the continuously level-dependent pair of generator and drift matrices. It has also been shown that this discretization approach is numerically efficient and stable due to the linear dependence of the overall computational complexity of the MRMFQ solver with respect to the number of regimes.

\section{The $MAP/M/s+G$ Queue and Its Stationary Solution}
\label{stationary}
We study the $MAP/M/s+G$ call center queueing model with $s$ servers, infinite waiting room, and generally distributed patience times. 
Calls (or customers) arrive at the system according to a Markovian Arrival Process (MAP) which is a versatile process that allows dependence in successive call interarrival times; see Lucantoni et al. \cite{lucantoni_adv90} for an elaborate discussion of MAPs.
A MAP with order $m$ is characterized with the matrix pair $(C,D)$, namely $MAP(C,D)$, both characterizing matrices being $m \times m$, $C=\{C_{ij}\}$ having negative
diagonal elements and non-negative off-diagonal elements, $D=\{D_{ij} \}$ being
non-negative, and $E = C+D$ is an irreducible
infinitesimal generator. The evolution of the MAP is described through the following.
Assume that the Markov process with
generator $E$ is in state $j$, $1 \leq j \leq m$. After an
exponentially distributed time with parameter $-C_{jj}$, a transition to another state $k \neq j$
occurs without an arrival with probability $\frac{C_{jk}}{-C_{jj}}$
or a transition to state $l$ occurs with an arrival with
probability $\frac{D_{jl}}{-C_{jj}}$. Let $\alpha$ be the
stationary probability vector of the phase process with generator $E$
so that $\alpha$ satisfies
\begin{equation}
\alpha E = 0, \; \alpha e= 1, \label{alpha}
\end{equation}
where $e$ denotes a vector of ones of appropriate size.
The mean arrival rate $\lambda$ for the $MAP(C,D)$ is then given by Lucantoni \cite{lucantoni_sm91}:
\begin{equation}
\lambda = \pi d^{\ast} = \pi D e, \label{lambda}
\end{equation}
with $d^{\ast}=\{ d_j ^{\ast}\}$ and $d_j ^{\ast}$ being the call arrival rate at state $j$.
Moreover, the embedded stationary probability vector (embedded at post-arrival epochs) denoted by $\pi$ satisfies
\begin{equation} 
\pi P=P, \; \pi e=1, \label{pi}
\end{equation}
where $P=(-C)^{-1}D$ denotes the $m \times m$ probability transition matrix at the embedded epochs of arrivals.
The relation between these two stationary vectors is given by
\begin{equation}
\alpha = \lambda \pi (-C)^{-1}.
\end{equation}

A subcase of MAP is phase-type (PH-type) which is a commonly used model
for independent and identically distributed 
non-exponential interarrival and/or service times; see Neuts \cite{neuts81}. To describe a
PH-type distribution, a Markov process is defined on the states
$\{1,2,\ldots,m,m+1\}$ with the absorbing state $m+1$, initial probability vector $(\beta,0)$, and an
infinitesimal generator \[ \left[ \begin{array}{cc}
B & B^0\\
0 & 0 \end{array}
\right],
\]
where $\beta$ is a row vector of size $m$, the subgenerator
$B$ is $m \times m$, and $B^0$ is a column vector of size $m$ such that $B^0=-Be$. The distribution of the time
till absorption into the absorbing state $m+1$, denoted by the random variable $X$, is called PH-type characterized with the pair $(\beta,B)$ or order $m$, i.e., $X \sim PH(\beta,B)$ where the notation $\sim$ is synonymous with \enquote{distributed according to}. A PH-type arrival process with representation
$(\beta,B)$ is indeed a MAP for which $C=B$ and $D = B^0 \beta$.
The cumulative distribution function (cdf) and probability density function (pdf) of $X \sim PH(\beta,B)$, denoted by $F_X(x)$ and $f_X(x)$, respectively, are given as: 
\begin{equation}
F_X(x)=1 -\beta e^{Bx} e, \;
f_X(x)=-\beta e^{Bx} B e, \mbox{~~~for~} x \geq 0.
\label{phdensity}
\end{equation}
A generalization of the PH-type distribution is the so-called Matrix Exponential (ME) distribution
Asmussen and Fackrell \cite{AsmussenBladt97}, 
Fackrell \cite{fackrell_thesis},
He and Zhang \cite{he_aap07}. We say $X \sim ME(\beta,B)$ with order $m$ when the pdf of the random variable $X$ is of the same form \eqref{phdensity} as in PH-type distributions, but the parameters $\beta$ and $B$ do not necessarily have a probabilistic interpretation pertaining to PH-type distributions. However, the ME-distributed variable $X$ still has a legitimate density, $f_X(x)\geq 0, \forall x\geq 0,\int_{0}^{\infty} \;f_X(x) dx=1$. Stochastic models relying on PH-type distributions with probabilistic matrix parameters are shown to be extendable to those with ME-distributions as well: see Bean and Nielsen  \cite{Bean2010},
Bucholtz and Telek \cite{[BUCH09a],[BUCH10b]}.

For the $MAP/M/s+G$ queueing system, there are $s$ statistically identical servers (or agents) to serve the incoming calls (customers). The service time of an individual customer is exponentially distributed with parameter $\mu$ and the system load is defined as $\rho = \frac{\lambda}{s \mu}$.
An incoming call is directed to one of idle servers randomly, i.e., zero wait. If all the servers are busy, then the customer joins a First-Come First-Served (FCFS) queue to wait with no limit on the waiting room capacity.
When waiting customers run out of patience before their service start, they abandon the queue. Mathematically, impatience occurs when the queue wait time exceeds
the abandonment time random variable $A$ which is generally distributed with pdf $f_A(\cdot)$. In Zelytn and Mandelbaum \cite{Zeltyn.2005}, various distributional choices for the random variable $A$ are studied through empirical call center data. 
More general PH-type distributions for service times are also possible to incorporate with the proposed matrix-analytic method but such extensions are left as future work.

In order to describe the $MAP/M/s+G$ queueing system, we first define the virtual waiting time $V(t)$ which stands for the waiting time of a virtual customer that would arrive at time $t$ provided this customer would eventually be served, i.e., successful. 
If there are strictly less than $s$ customers in the system, then $V(t)=0$. If there are exactly $s$ customers in the system, i.e., all servers are busy and the queue is empty, $V(t)$ amounts to the time required until the first customer departure.
Generally, when the number of customers in the system is larger than or equal to $s$,
then $V(t)=x > 0$ and a virtual incoming customer at time $t$ would decide to wait in the queue if $A > x$ for eventual successful service or equivalently with probability 
\(
g_s(x)=\int_{0}^{x}f_A(y)\:dy.
\)
Similarly, an incoming virtual customer at time $t$ would be tagged as \enquote{abandoned} at the epoch of arrival which would occur with probability 
\(
g_a(x)=1 - g_s(x).
\)
A tagged customer will wait in the queue but will surely abandon before the service starts and its queue wait time will not have any effect on that of the successful calls, i.e., non-abandoning calls. Therefore, in our model, $V(t)$ keeps track of successful customers' waiting times only. 

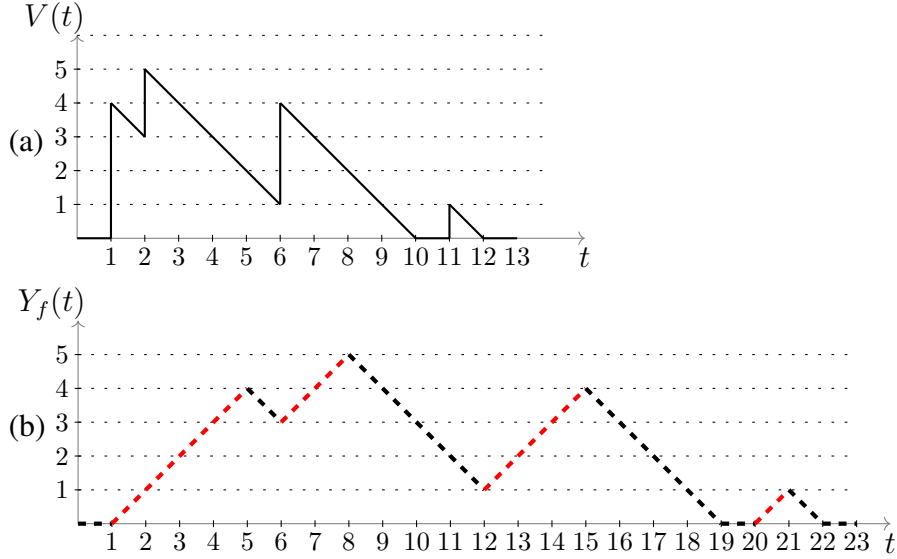
\begin{figure}[t]
	\begin{tikzpicture}[scale=0.45]
	\draw [gray, <->] (0,6) -- (0,0) -- (15,0);
	
	\node [below] at (15,0){\large $t$} ;
	\node [left] at (0.5,6.5) {\large $V(t)$};
	\node[align=center, below] at (-1.5,3.5)%
	{\large (a)};
	
	\draw[dash pattern=on 1pt off 4pt] (0,1) -- (14,1);
	\draw[dash pattern=on 1pt off 4pt] (0,2) -- (14,2);
	\draw[dash pattern=on 1pt off 4pt] (0,3) -- (14,3);
	\draw[dash pattern=on 1pt off 4pt] (0,4) -- (14,4);
	\draw[dash pattern=on 1pt off 4pt] (0,5) -- (14,5);
	\draw[dash pattern=on 1pt off 4pt] (0,6) -- (14,6);
	
	\draw (1,-0.1) -- (1,0.1);
	\draw (2,-0.1) -- (2,0.1);
	\draw (3,-0.1) -- (3,0.1);
	\draw (4,-0.1) -- (4,0.1);
	\draw (5,-0.1) -- (5,0.1);
	\draw (6,-0.1) -- (6,0.1);
	\draw (7,-0.1) -- (7,0.1);
	\draw (8,-0.1) -- (8,0.1);
	\draw (9,-0.1) -- (9,0.1);
	\draw (10,-0.1) -- (10,0.1);
	\draw (11,-0.1) -- (11,0.1);
	\draw (12,-0.1) -- (12,0.1);
	
	\draw (-0.1,1) -- (0.1,1);
	\draw (-0.1,2) -- (0.1,2);
	\draw (-0.1,3) -- (0.1,3);
	\draw (-0.1,4) -- (0.1,4);
	\draw (-0.1,5) -- (0.1,5);

	\node[align=center, left] at (0,1)%
	{$1$};
	\node[align=center, left] at (0,2)%
	{$2$};
	\node[align=center, left] at (0,3)%
	{$3$};
	\node[align=center, left] at (0,4)%
	{$4$};
	\node[align=center, left] at (0,5)%
	{$5$};

	\node[align=center, below] at (1,0)%
	{$1$};
	\node[align=center, below] at (2,0)%
	{$2$};
	\node[align=center, below] at (3,0)%
	{$3$};
	\node[align=center, below] at (4,0)%
	{$4$};
	\node[align=center, below] at (5,0)%
	{$5$};
	\node[align=center, below] at (6,0)%
	{$6$};
	\node[align=center, below] at (7,0)%
	{$7$};
	\node[align=center, below] at (8,0)%
	{$8$};
	\node[align=center, below] at (9,0)%
	{$9$};
	\node[align=center, below] at (10,0)%
	{$10$};
	\node[align=center, below] at (11,0)%
	{$11$};
	\node[align=center, below] at (12,0)%
	{$12$};
	\node[align=center, below] at (13,0)%
	{$13$};
	
	\draw[thick,black] (0,0) -- (1,0);
	\draw[thick,black] (1,0) -- (1,4);
	\draw[thick,black] (1,4) -- (2,3);
	\draw[thick,black] (2,3) -- (2,5);
	\draw[ thick,black] (2,5) -- (6,1);
	\draw[ thick,black] (6,1) -- (6,4);
	\draw[thick,black] (6,4) -- (10,0);
	\draw[ thick,black] (10,0) -- (11,0);
	\draw[ thick,black] (11,0) -- (11,1);
	\draw[thick,black] (11,1) -- (12,0);
	\draw[ thick,black] (12,0) -- (13,0);
	\end{tikzpicture}
	\newline
	\begin{tikzpicture}[scale=0.45]
	
	\draw [gray, <->] (0,6) -- (0,0) -- (24,0);
	
	\node [below] at (24,0){\large $t$} ;
	\node [left] at (0.5,6.5) {\large $Y_f(t)$};
	\node[align=center, below] at (-1.5,3.5)%
	{ \large (b)};
	
	\draw[dash pattern=on 1pt off 4pt] (0,1) -- (23,1);
	\draw[dash pattern=on 1pt off 4pt] (0,2) -- (23,2);
	\draw[dash pattern=on 1pt off 4pt] (0,3) -- (23,3);
	\draw[dash pattern=on 1pt off 4pt] (0,4) -- (23,4);
	\draw[dash pattern=on 1pt off 4pt] (0,5) -- (23,5);
	
	\draw (1,-0.1) -- (1,0.1);
	\draw (2,-0.1) -- (2,0.1);
	\draw (3,-0.1) -- (3,0.1);
	\draw (4,-0.1) -- (4,0.1);
	\draw (5,-0.1) -- (5,0.1);
	\draw (6,-0.1) -- (6,0.1);
	\draw (7,-0.1) -- (7,0.1);
	\draw (8,-0.1) -- (8,0.1);
	\draw (9,-0.1) -- (9,0.1);
	\draw (10,-0.1) -- (10,0.1);
	\draw (11,-0.1) -- (11,0.1);
	\draw (12,-0.1) -- (12,0.1);
	\draw (13,-0.1) -- (13,0.1);
	\draw (14,-0.1) -- (14,0.1);
	\draw (15,-0.1) -- (15,0.1);
	\draw (16,-0.1) -- (16,0.1);
	\draw (17,-0.1) -- (17,0.1);
	\draw (18,-0.1) -- (18,0.1);
	\draw (19,-0.1) -- (19,0.1);
	\draw (20,-0.1) -- (20,0.1);
	\draw (21,-0.1) -- (21,0.1);
	\draw (22,-0.1) -- (22,0.1);
	\draw (23,-0.1) -- (23,0.1);
	
	\draw (-0.1,1) -- (0.1,1);
	\draw (-0.1,2) -- (0.1,2);
	\draw (-0.1,3) -- (0.1,3);
	\draw (-0.1,4) -- (0.1,4);
	\draw (-0.1,5) -- (0.1,5);

	\node[align=center, left] at (0,1)%
	{$1$};
	\node[align=center, left] at (0,2)%
	{$2$};
	\node[align=center, left] at (0,3)%
	{$3$};
	\node[align=center, left] at (0,4)%
	{$4$};
	\node[align=center, left] at (0,5)%
	{$5$};

	\node[align=center, below] at (1,0)%
	{$1$};
	\node[align=center, below] at (2,0)%
	{$2$};
	\node[align=center, below] at (3,0)%
	{$3$};
	\node[align=center, below] at (4,0)%
	{$4$};
	\node[align=center, below] at (5,0)%
	{$5$};
	\node[align=center, below] at (6,0)%
	{$6$};
	\node[align=center, below] at (7,0)%
	{$7$};
	\node[align=center, below] at (8,0)%
	{$8$};
	\node[align=center, below] at (9,0)%
	{$9$};
	\node[align=center, below] at (10,0)%
	{$10$};
	\node[align=center, below] at (11,0)%
	{$11$};
	\node[align=center, below] at (12,0)%
	{$12$};
	\node[align=center, below] at (13,0)%
	{$13$};
	\node[align=center, below] at (14,0)%
	{$14$};
	\node[align=center, below] at (15,0)%
	{$15$};
	\node[align=center, below] at (16,0)%
	{$16$};
	\node[align=center, below] at (17,0)%
	{$17$};
	\node[align=center, below] at (18,0)%
	{$18$};
	\node[align=center, below] at (19,0)%
	{$19$};
	\node[align=center, below] at (20,0)%
	{$20$};
	\node[align=center, below] at (21,0)%
	{$21$};
	\node[align=center, below] at (22,0)%
	{$22$};
	\node[align=center, below] at (23,0)%
	{$23$};
	
	\draw[ultra thick,dashed,black] (0,0) -- (1,0);
	\draw[ultra thick,dashed,red] (1,0) -- (5,4);
	\draw[ultra thick,dashed,black] (5,4) -- (6,3);
	\draw[ultra thick,dashed,red] (6,3) -- (8,5);
	\draw[ultra thick,dashed,black] (8,5) -- (12,1);
	\draw[ultra thick,dashed,red] (12,1) -- (15,4);
	\draw[ultra thick,dashed,black] (15,4) -- (19,0);
	\draw[ultra thick,dashed,black] (19,0) -- (20,0);
	\draw[ultra thick,dashed,red] (20,0) -- (21,1);
	\draw[ultra thick,dashed,black] (21,1) -- (22,0);
	\draw[ultra thick,dashed,black] (22,0) -- (23,0);
	\end{tikzpicture}
	\caption{Sample paths of (a) the virtual waiting time process $V(t)$ and (b) auxiliary random process $Y_f(t)$.\label{fig:samplePath}}
\end{figure}

The sample path for the process $V(t)$ is given in Fig.~\ref{fig:samplePath}(a) for an example scenario with three servers. In this example, we have customers numbered $
1,\ldots,7$ arriving at instances $0, 0, 1, 2, 4, 6, 11$, with service times $5,7,9,7,\times,8,5$, respectively. At $t=1^-$, $V(1^-)=0$, customers 1 and 2 are in service with remaining service times of 4 and 6, respectively, and at $t=1$, the customer 3 arrives with a service time of 9.
Obviously, customer 3 will have a zero queue wait and $V(1^+)=4$ which turns out to be the minimum of the three remaining service times. 
At $t=2^-$, $V(2^-)=3$, and the customers 1, 2, and 3 are in service with remaining times 3, 5, and 8, respectively, and the customer 4 arrives with a service time of 7 which cannot be served immediately but instead begins to wait in the queue.
At $t=2^+$, $V(2^+)=5$ which turns out to be time until the departure of 
customer 2. At $t=4$, customer 5 arrives when $V(4^-)=2$ but it is tagged as \enquote{abandoned} since the its queue wait turns out to be larger than a realization of the random variable $A$. 
Therefore, this customer will not have an effect on $V(t)$ at $t=4$ and its service time is thus represented by a don't care $\times$.
At $t=5$, customer 1 leaves and customer 4 starts service and subsequently
$V(t)$ evolves in Fig.~\ref{fig:samplePath}(a).
Due to abrupt jumps, the virtual waiting time process can not be modeled as an ordinary Markov fluid queue. Therefore, we introduce an auxiliary random process $Y_f(t)$ by replacing the abrupt jumps in the original process $V(t)$ with linear increases with a drift of $+1$ as illustrated Fig.~\ref{fig:samplePath}(b) with red dashed lines. Moreover, it is clear from sample path arguments that the steady-state distribution of the process $V(t)$ can be derived from that of $Y_f(t)$ by censoring out the states corresponding to a drift of $+1$.


Next we focus on the CFFQ model for the fluid process $Y_f(t)$. For this purpose, we first define the two-dimensional process $Y_m(t)= \{ S(t),M(t);t\geq 0 \}$ where $S(t) \in {\mathcal S} = \{0,1,\ldots, s-2,s-1,s \}$ stands for the number of busy servers at time $t$ with two exceptions: the state $s-1$ refers to the part of the curve in Fig.~\ref{fig:samplePath}(b) where the virtual waiting time is strictly decreasing when $Y_f(t)>0$ and also $s-1$ busy servers and $Y_f(t)=0$; and state $s$ refers to the part of the same curve where $Y_f(t)$ is strictly increasing. Moreover, $M(t) \in {\mathcal M}= \{1,2,\dots,m \}$ keeps track of the MAP state at time $t$. With this definition, the joint process 
${\mathbf Y(t)}= (Y_f(t),Y_m(t))$, $t\geq 0,$ where $0\leq Y_f(t) < \infty$ is a CFFQ process characterized with the matrix pair $(Q_Y(x),R_Y(x))$ which are provided below on the basis of the enumeration of the joint state of $Y_m(t)$ as $(0,1),(0,2),\ldots,(0,m),(1,0),\ldots,(1,m),\ldots,(s,m)$:
\begin{align}
Q_Y(0) &= \left[ \begin{array}{c|c|c|c|c|c}
C & D & & & & \\ \hline
\mu I_m & C-\mu I_m & D & & & \\ \hline
& 2\mu I_m & C-2 \mu I_m & D & & \\ \hline
& & \ddots & \ddots & \ddots & \\ \hline
& & & (s-1) \mu I_m & C-(s-1) \mu I_m & D \\ \hline
& & & & 0_m & 0_m 
\end{array} \right],\label{QY0} \\
R_Y(0) &= {\bf diag} \{0_{ms},I_m \},  \label{RY0}
\end{align}
and for $x>0$:
\begin{align}
Q_Y(x) &= \left[ \begin{array}{c|c|c}
0_{m(s-1)} & 0_m & 0_m\\ \hline
0_{m \times m(s-1)} & C + Dg_a(x)  & D g_s(x) \\ \hline
0_{m \times m(s-1)}& s \mu I_m & -s \mu I_m  
\end{array} \right],\label{QYx} \\
R_Y(x) &= {\bf diag} \{-I_{ms},I_m \}, \label{RYx} 
\end{align}
where $0_m$ and $I_m$ stand for a square matrix of zeros and the identity matrix of size $m$, respectively, $0_{m \times n}$ stands for a matrix of zeros of size $m \times n$, and ${\bf diag}$ operator stands for the diagonal concatenation of its matrix input arguments:
\[
{\bf diag}\{A_1,A_2,\ldots,A_l \} = \left[ \begin{array}{cccc}
A_1 & 0 & \cdots & 0 \\
0 & A_2 & \cdots & 0 \\
\vdots & \vdots & \ddots & 0 \\
0 & 0 & \cdots & A_l 
\end{array}
\right],
\]
for diagonal $A_j,1 \leq j \leq l.$

If the random variable $A$ is such that the abandonment function $g_a(x)$
is originally piece-wise constant, then the process ${\mathbf Y(t)}$ is an MRMFQ and its steady-state solution can be obtained through the method outlined in Section~\ref{MRMFQ}. If $A$ is a continuous random variable, then the CFFQ can approximately be reduced to a $K$-regime MRMFQ with the following proposed discretization approach.
We first choose the boundary points $T^{(k)}$ of the MRMFQ such that 
$g_a(T^{(k)})=\frac{k}{K},k=1,2,\ldots,K-1$, and we approximate $g_a(x)$ in regime $k, 1 \leq k < K$, by $g_a \left(\frac{1}{2} (T^{(k-1)}+T^{(k)}) \right)$.
Only for the rightmost regime $K$, we approximate $g_a(x)$ by $g_a(\infty)=1$. If $A$ is a hybrid random variable, then these two methods can be combined, which we will not delve into, for the numerical examples of the the current paper. 

We will now focus on the steady-state solution of the MRMFQ ${\mathbf Y(t)}$. Since the non-zero drift matrices are identical for each regime and boundary for $x>0$, there can not be a probability mass accumulation at a boundary point other than the origin. It is also clear that there can not be any probability mass also at the origin for states $(\cdot,s)$ since the drift rate at these states are positive. Moreover, for the states $(i,\cdot)$ where $i \in \{0,1,\ldots,s-2 \}$, we can not have any probability density for $x>0$. In light of these observations, let $c^{(0)}_{(i,j)}$ denote the probability mass at the origin for the state $(i,j)$ where $i \in \{0,1,\ldots,s-1\}$ and $j \in \{1,2,\ldots,m \}$ and also let $f_{(i,j)}(x), x>0$ denote the steady-state joint pdf for the state $(i,j)$ where $i \in \{s-1,s\}$ and $j \in \{1,2,\ldots,m \}$ for the MRMFQ ${\mathbf Y(t)}$. Since the state $s$ corresponding to abrupt jumps in Fig.~\ref{fig:samplePath}(a) needs to be censored out, we need to perform the following normalization for the censoring operation:
\begin{equation}
\sum_{i=0}^{s-1} \sum_{j=1}^m c^{(0)}_{(i,j)} + \sum_{j=1}^m \int_{x=0}^{\infty} 
f_{(s-1,j)}(x)\:dx =1.
\end{equation}
After this normalization, $c^{(0)}_{(i,j)}$ and $f_{(i,j)}(x)$ are now associated with 
the virtual waiting time process $V(t)$ while holding on to the same definition.
Let $W$ denote the steady-state queue wait time of customers. The first performance metric of interest is the probability of zero wait $\Pr\{ W=0 \}$:
\begin{equation}
\Pr \{ W=0 \}=\frac{1}{\lambda} \sum_{i=0}^{s-1} \sum_{j=1}^m  c^{(0)}_{(i,j)} d_j^{\ast}.
\end{equation}
Let $\Pr \{\mathcal{A}\}$ denote the probability of abandonment for which we provide the following expression:
\begin{equation}
\Pr \{ \mathcal{A}\} =  \frac{1}{\lambda} \sum_{j=1}^m d_j^{\ast} \left( \int_{x=0}^{\infty} 
f_{(s-1,j)}(x) g_a(x) \:dx \right)  
\end{equation}
The probability of success denoted by $\Pr \{ \mathcal{S} \} =  1 - \Pr \{ \mathcal{A} \}$ can also be given by the following expression:
\begin{equation}
\Pr \{ \mathcal{S} \}  =  \Pr \{ \mathcal{S},W>0 \} + \Pr \{ W=0 \},
\end{equation}
where
\begin{equation}
\Pr \{ \mathcal{S},W>0 \} =  \frac{1}{\lambda}  \sum_{j=1}^m d_j^{\ast}\left( \int_{x=0}^{\infty} 
f_{(s-1,j)}(x) g_s(x) \:dx \right)   
\end{equation}
denotes the probability of a non-zero wait successful customer. 
Let $f_{W|\mathcal{S},W>0}(x)$ denote the waiting time density conditioned on non-zero wait successful customers which is given by the following expression:
\begin{equation}
f_{W|\mathcal{S},W>0}(x)=\frac{1}{\lambda \Pr \{ \mathcal{S},W>0 \}} \sum_{j=1}^m d_j^{\ast} 
f_{(s-1,j)}(x) g_s(x) .
\end{equation}
We also let  \begin{equation}
F_{W|\mathcal{S},W>0}(x) =  \int_0^{x }f_{W|\mathcal{S},W>0}(\tau) \:d\tau,
\end{equation} 
be the corresponding conditional cdf of the waiting time of non-zero wait successful customers.

\section{Analysis of First Passage Times for the $MAP/M/s+G$ Queue}
\label{firstpassagesection}
\subsection{First Passage Time for the Virtual Waiting Time}
 First, we are interested in the first passage time distribution of the virtual waiting time for the $MAP/M/s+G$ queue.
For this purpose, let $\kappa(a,b,\pi_0,\theta_0)$ denote the first passage time from level $a$ to level $b$ for the virtual waiting time $V(t)$:
\begin{equation}
\kappa(a,b,\pi_0,\theta_0) = \inf_t \{ V(t)=b \: | \: V(0)=a,S(0) \sim \pi_0 , M(0) \sim \theta_0\}.
\label{firstpassagecdf}
\end{equation}
In the above definition, the notation $\sim$ is used to denote the following:
\begin{equation}
\pi_0(i)=\Pr \{ S(0) = i\}, i=0,\ldots,s, \; \theta_0(j)=\Pr \{ M(0) = j\} ,j=1,\ldots,m. 
\end{equation}
Obviously, it should hold that $\Pr \{ S(0) = s\}=0$, and $\Pr \{ S(0) = s-1\}=1$ when $a>0$. The following probability is of interest for the virtual waiting time in the $MAP/M/s+G$ queue:
\begin{equation}
F_v^{a,b,\pi_0,\theta_0}(\tau)  = \Pr \{\kappa(a,b,\pi_0,\theta_0) < \tau \}.
\label{firstpassagedistribution}
\end{equation}
This probability is indicative of whether a certain level for the virtual waiting time is to be reached within a time horizon $\tau$ and can potentially be used in risk analysis.


In this paper, a random variable $U$ is used for approximating the deterministic time horizon, $\tau$, which is either PH-type or ME-distributed with mean $\tau$ and order $\ell$. The most well-known method is to use the so-called Erlang-$\ell$ distribution of order $\ell$ which is PH-type, characterized with the matrix pair $(\beta_{\ell},B_{\ell})$ where
\[ 
\beta_\ell{} = \left[ \begin{array}{cccc}  1 & 0 & \cdots & 0 \end{array} \right], 
B_{\ell}= \frac{\ell}{\tau} \begin{bmatrix}
-1 & 1 & & & \\
& \ddots & \ddots & \\
& & -1 & 1 \\
& & & -1
\end{bmatrix}. \]
The Squared Coefficient of Variation (SCV) of the Erlang-$\ell$ distribution is $1/\ell$ and as $\ell \rightarrow \infty$, one has the least variable PH-type approximation to the deterministic variable $\tau$. However, a family of ME distributions of order $\ell$, called the concentrated ME distribution CME-$\ell$, are recently proposed by Horvath et al. \cite{cme} which are shown to achieve a SCV much lower than $1/\ell$ and the asymptotic behavior of the SCV of the proposed CME distribution of order $\ell$ is actually $2/\ell^2$. In \cite{cme}, an algorithm is presented to construct a CME-$\ell$ distribution for $\ell \leq 47$, but in more recent work by Horvath, Horvath, and Telek, ite{cme2}, numerically stable and efficient algorithms are presented for higher orders for odd values of $\ell$.
We study both Erlang-$\ell$ and CME-$\ell$ approximations in the numerical examples of this paper, the latter being based on the algorithm proposed in \cite{cme2}. In either case, we start with $U \sim PH(\beta,B)$ for the development of the CFFQ model needed for the first passage times. The algorithm for the case $U \sim ME(\beta,B)$ will be identical.

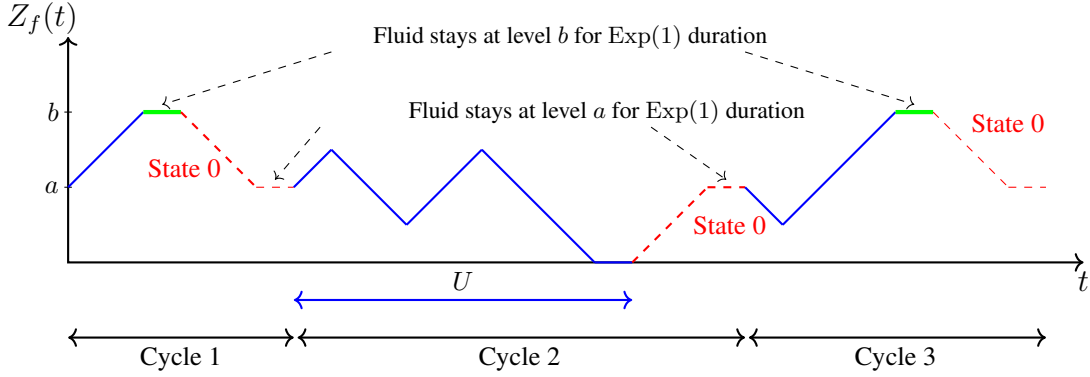
\begin{figure}[tb]
		\centering
	\begin{tikzpicture}[scale=0.5]
	
	\draw [thick, <->] (0,6) -- (0,0) -- (27,0);

	\node [below] at (27,0){\large $t$} ;
	\node [left] at (0.5,6.5) {\large $Z_f(t)$};
	\node[align=center, left] at (0,2)%
	{$a$};
	\node[align=center, left] at (0,4)%
	{$b$};
	
	\draw (-0.1,2) -- (0.1,2);
	\draw (-0.1,4) -- (0.1,4);

	\draw [thick,blue] (0,2) -- (2,4);
	\draw [ultra thick,green] (2,4) -- (3,4);
	\draw [dashed,thick,red] (3,4) -- (5,2);
	\draw [dashed,red] (5,2) -- (6,2);
	\draw [thick,blue] (6,2) -- (7,3);
	\draw [thick,blue] (7,3) -- (9,1);
	\draw [thick,blue] (9,1) -- (11,3);
	\draw [thick,blue] (11,3) -- (14,0);
	\draw [thick,blue] (14,0) -- (15,0);
	\draw [dashed,thick,red] (15,0) -- (17,2);
	\draw [dashed,thick,red] (17,2) -- (18,2);
	\draw [thick,blue]       (18,2) -- (19,1);
	\draw [thick,blue]       (19,1) -- (22,4); 
	\draw [ultra thick,green] (22,4) -- (23,4); 
	\draw [dashed,red]        (23,4) -- (25,2); 
	\draw [dashed,red]        (25,2) -- (26,2); 
	
	\draw [thick,<->] (0,-2) -- (6,-2);
	\node[align=center,below] at (3,-2)%
	{Cycle 1};
	\draw [thick,<->] (6.1,-2) -- (18,-2);
	\node[align=center,below] at (12,-2)%
	{Cycle 2}; 
	\draw [thick,<->] (18.1,-2) -- (26,-2);
	\node[align=center,below] at (22,-2)%
	{Cycle 3};
	\draw [thick,<->,blue] (6,-1) -- (15,-1);
	\node[align=center,below] at (10.5,0)%
	{$U$};
	
	\node[align=center,left] at (4.3,2.5)%
	{\color{red} State 0};
	\node[align=center,left] at (18.8,1)%
	{\color{red} State 0};
	\node[align=center,left] at (26.2,3.7)%
	{\color{red} State 0};
	
	\draw [dashed,->] (15.5,3.6)  --(17.5,2.2);
	\draw [dashed,->] (6.7,3.6)  --(5.5,2.2);
	\node[align=center,left] at (20,4)%
	{\small Fluid stays at level $a$ for $\rm{Exp} (1)$ duration };

	\draw [dashed,->] (7,5.6) -- (2.5,4.2);
	\draw [dashed,->] (18,5.6) -- (22.5,4.2);
	\node[align=center,left] at (19,6)%
	{\small Fluid stays at level $b$ for $\rm{Exp} (1)$ duration };
	\end{tikzpicture}
	\caption{Sample path of the random process $Z_f(t), t\geq 0$.\label{fig:samplePath2}}
\end{figure}

%
For the purpose of obtaining the first passage time distribution in (\ref{firstpassagedistribution}), we proceed by introducing an auxiliary CFFQ $\mathbf{Z(t)}=(Z_f(t),Z_m(t)), t \geq 0$.
In this case, we let the modulating process $Z_m(t)$ to have the following state space:
\begin{equation} 
Z_{m}(t) \in \mathcal{Z} = \{0\} \cup \left( \mathcal{U} \otimes \mathcal{S} \otimes 
\mathcal{M} \right)  , \quad \mathcal{U}=\{1,2,\ldots,\ell \}.
\end{equation}
The size of the state space for $Z_{m}(t)$ is $1+\ell m (s+1)$ and $\ell m (s+1)$ of these states correspond to the triple $(\sigma, i,j)$, $1 \leq \sigma \leq \ell$, $0\leq i \leq s$, and $1 \leq j\leq m$, $\sigma$ keeping track of the phase of the time horizon $U$ whereas 
the pair $(i,j)$ representing the phase of two-dimensional modulating process $Y_m(t)= \{ S(t),M(t) \}$ which was already used to model the steady-state behavior of the $MAP/M/s+G$ queue in Section~\ref{stationary}. State $0$ is an auxiliary state used to reset
the process $Z_f(t)$ to the initial conditions when the cycle termination conditions are met and prepare the fluid level and the modulating processes for the subsequent cycle. The fluid level process $Z_f(t)$ is a finite capacity continuous-valued process in the interval $[0,b]$ if $a<b$ or $[b,\infty)$ if $b<a$. A sample path of $Z_f(t)$ is depicted in Fig.~\ref{fig:samplePath2} for a span of three consecutive cycles. The fluid level process starts each cycle at level $a$ and
evolves according to the original fluid process $Y_f(t)$ and to its generator, and rate matrices, similar to the illustration in Fig.~\ref{fig:samplePath}(b). In order to keep track of time, a timer is started at the beginning of each cycle which would expire when the absorbing state of $U$ is reached. Until then, the timer evolves according to the generator $B$. Either when the fluid level reaches $b$ before the timer expires (cycles 1 and 3 in Fig.~\ref{fig:samplePath2}), or when the timer expires without the fluid level reaching $b$ (cycle 2), the auxiliary state $0$ is entered. In either case, we force the fluid level back to the initial value $a$ and reset the states (state of the timer included) according to $\pi_0$, $\theta_0$, and $\beta,$ in order to start a new cycle. Subsequently, this pattern repeats forever.

Note that for the purpose of keeping the time appropriately, when in state $(\cdot,s,\cdot)$, time needs to be frozen since jumps at those states are abrupt in reality (see Fig.~\ref{fig:samplePath}(a)). This corresponds to stopping the timer introduced above at states $(\cdot,s,\cdot)$ by blocking any transition from $(\sigma_1,s,\cdot)$ to $(\sigma_2,s,\cdot)$ when $\sigma_1 \neq \sigma_2$ and also to state $0$.
At the end of each cycle, the fluid level stays at state $0$ and level $a$ for an exponentially distributed duration of time with an arbitrary parameter (we use the unit parameter in this paper) before a new cycle can start. Similarly, when the level $b$ is reached before the timer expires, the fluid level stays at this level for an exponentially distributed duration with unit parameter after which the level is forced to reach level $a$. This cycle termination pattern enables us to acquire the ratio of cycles that terminate due to a hit to level $b$, to all cycles. Eventually, with the infinitely many repetitions of cycles, this ratio will lead us to the first passage probability within the given time horizon. 

The constructed CFFQ provides a single trajectory of infinitely many cycles and the steady-state solution of this cyclic CFFQ will then provide an approximation for the first passage time distribution of the virtual waiting time evaluated at a given point $\tau$ in Eqn.~(\ref{firstpassagedistribution}). This approximation will be numerically shown to converge to the exact results as $\ell \rightarrow \infty$ for piece-wise constant $g_a(x)$ case. In case of continuous/hybrid $g_a(x)$, we further need the condition $K \rightarrow \infty$ for numerical convergence since the CFFQ will approximately be reduced to a $K$-regime MRMFQ with the same discretization approach followed in Section~\ref{stationary}.

By the ordering of the states of $Z_m(t)$ as 
\[ 0,(1,0,1),(1,0,2),\ldots,(1,0,m),(1,1,0),\ldots,(\ell,s,m),
\]
the expressions for the matrix pair $\left( Q_Z(x),R_Z(x)\right) $ characterizing the CFFQ process $\mathbf{Z(t)}$ can now be given. For this purpose, we first need to define $\tilde{I}$ which is an identity matrix of size $m(s+1)$ except for a zero matrix of size $m \times m$ in the south-east corner:
\begin{equation}
\tilde{I} ={\bf diag}\{I_{ms},0_m\},
\label{newdiag}
\end{equation}
and also a column vector $\tilde{e}$ of ones of size $m(s+1)$ except for its last $m$ entries which are all zeros: 
\begin{equation}
\tilde{e}^T =\{1_{1 \times ms}, 0_{1 \times m}\}.
\label{ee}
\end{equation}
Zero entries in the matrices $\tilde{I}$ and $\tilde{e}$ are key for freezing the timer when in state $(\cdot,s,\cdot)$ by setting the transition rate from $(\sigma_1,s,\cdot)$ to $(\sigma_2,s,\cdot)$ and to state $0$ as zero whenever $\sigma_1 \neq \sigma_2$.

Subsequently, the matrix $Q_Z(x)$ is written as:
\begin{eqnarray}
Q_Z(b) & = &\left[ \begin{array}{c|c}
0 & 0 \\ \hline
e & -I_{\ell m(s+1)} 
\end{array} \right],
\label{Q_ZB} \\
Q_Z(x) & = & \left[ \begin{array}{c|c}
-h(x) &  h(x)(\alpha \otimes \pi_0 \otimes \theta_0) \\ \hline
B^0 \otimes \tilde{e} & I_{\ell} \otimes Q_Y(x) + B \otimes \tilde{I}
\end{array} \right], \;{\rm for } \;x \neq b, 
\label{Q_YMAP}
\end{eqnarray}
where
\begin{equation}
h(x)=0 \; {\rm for} \; x\neq a, \; h(a)=1.
\end{equation}
When $a > 0$ and $x \neq b$, the drift matrix $R_Z(x)$ is written as:
\begin{equation}
R_Z(x) = 
\begin{cases} 
{\bf diag}\{1,I_{\ell} \otimes R_Y(x)\}, & \text{if } x < a, \\
{\bf diag}\{-1,I_{\ell}\otimes R_Y(x)\}, & \text{if } x > a,   \\
{\bf diag} \{0,I_{\ell}\otimes R_Y(a)\}, & \text{if } x = a. \\
\end{cases}
\label{R_Z}
\end{equation}
On the other hand, if $a=0$ and when $x \neq b$, Eqn.~(\ref{R_Z}) requires a slight modification:
\begin{equation}
R_Z(x) = 
\begin{cases} 
{\bf diag}\{-1,I_{\ell}\otimes R_Y(x)\}, & \text{if } x > 0, \\
{\bf diag} \{0,I_{\ell}\otimes R_Y(0)\}, & \text{if } x = 0. \\
\end{cases}
\label{R_Z2}
\end{equation}
When $x=b$, we have:
\begin{equation}
R_Z(b) = 
\begin{cases} 
{\bf diag}\{-1,{\bf 0}_{{\ell }m(s+1)}\}, & \text{if } a < b,\\
{\bf diag}\{1,{\bf 0}_{{\ell}m(s+1)}\}, & \text{if } a > b, 
\end{cases}
\label{R_Z4}
\end{equation}
The identities given in (\ref{Q_ZB})-(\ref{Q_YMAP}) can be described as follows. The first block row of $Q_Z(x)$ ensures that the process $Z_m(t)$ remains at state $0$ as long as the fluid level is different from level $a$. The second block row of (\ref{Q_YMAP}) makes $Z_m(t)$ follow the behavior of $Y_m(t)$ and transition to state $0$ (cycle termination) may occur with either the expiration of the $U$-long time interval or with the fluid level hitting $b$. If the termination occurs with completion of the $U$-long interval (indicated by a state transition with rate $B^0$), $Z_m(t)$ moves to state $0$. Likewise, if the termination occurs by the fluid process hitting level $b$, the choice of $Q_Z(b)$ gives rise to $Z_m(t)$ making a transition to state $0$ after staying at level $b$ for an exponentially distributed duration with unit mean. The only case when $Z_m(t)$ can leave state $0$ is at fluid level $a$ according to the first block row of $Q_Z(x)$ in (\ref{Q_YMAP}) and the exit rate is $1$. Leaving state $0$, $Z_m(t)$ starts according to the initial distribution of the timer $U$, $\beta$, in a state representing $S(t) \sim \pi_0 , M(t) \sim \theta_0$. The interpretation of $R_Z(x)$ follows the same lines.

The above expressions complete the matrix representation of the CFFQ $\mathbf{Z(t)}$ whose steady-state solution can be found using the techniques described in Section~2. 
Let $c(b)$ denote the overall steady-state probability mass at level $b$ involving all the states $(\sigma,i,j)$, for $1 \leq \sigma \leq \ell$, $0\leq i \leq s$, $1 \leq j\leq m$ for the auxiliary MFQ ${\mathbf Z(t)}$. 
Also let $c_0(a)$ be the steady-state probability mass at level $a$ for state $0$. At each cycle of the trajectory for $Z_f(t)$, we end up at level $a$ during a visit to state $0$ for an exponentially distributed duration with unit parameter. For those cycles that the level $b$ is reached before the timer expires, we stay at level $b$ again for an exponentially distributed amount of time with unit parameter. 
Based on these observations, we have 
\begin{equation}
F_v^{a,b,\pi_0,\theta_0}(\tau)  \approx  \frac{c(b)}{c_0(a)} .
\label{firstpassage}
\end{equation}

%
\subsection{First Passage Time for the Actual Waiting Time}

Secondly, we are interested in the first passage time distribution of the actual waiting time for the $MAP/M/s+G$ queue.
For this purpose, let $\tilde{\kappa}(a,b,\pi_0,\theta_0)$ denote the first passage time from level $a$ to level $b$ for the actual waiting time:
\begin{equation}
\tilde{\kappa}(a,b,\pi_0,\theta_0) = \inf_t \{V(t^-)\geq b, \text{a successful arrival occurs at time } t \;| \;V(0)=a,S(0) \sim \pi_0 , M(0) \sim \theta_0\}.
\label{realfirstpassagecdf}
\end{equation}

\begin{figure}[tb]
	\begin{tikzpicture}[scale=0.4]
	
	\draw [thick, <->] (0,8) -- (0,0) -- (33,0);

	\node [below] at (33,0){\large $t$} ;
	\node [left] at (0.5,8.5) {\large $Z_f(t)$};
	\node[align=center, left] at (0,2)%
	{$a$};
	\node[align=center, left] at (0,4)%
	{$b$};
	
	\draw (-0.1,2) -- (0.1,2);
	\draw (-0.1,4) -- (0.1,4);

	\draw [thick,blue] (0,2) -- (4,6);
	\draw [thick,blue] (4,6) -- (5,5);
	\draw [dashed,thick,green] (5,5) -- (8,2);
	\draw [dashed,thick,green] (8,2) -- (9,2);
	\draw [thick,blue] (9,2) -- (10,3);
	\draw [thick,blue] (10,3) -- (12,1);
	\draw [thick,blue] (12,1) -- (15,5);
	\draw [thick,blue] (15,5) -- (17,3);
	\draw [thick,blue] (17,3) -- (18,4);
	\draw [dashed,thick,red] (18,4) -- (20,2);
	\draw [dashed,thick,red] (20,2) -- (21,2);
	\draw [thick,blue]       (21,2) -- (22,1);
	\draw [thick,blue]       (22,1) -- (27,6); 
	\draw [thick,blue] (27,6) -- (28,5); 
	\draw [dashed,thick,green]        (28,5) -- (31,2); 
	\draw [dashed,thick,green]        (31,2) -- (32,2); 
	
	\draw [thick,<->] (0,-2) -- (9,-2);
	\node[align=center,below] at (4.5,-2)%
	{Cycle 1};
	\draw [thick,<->] (9.1,-2) -- (21,-2);
	\node[align=center,below] at (15,-2)%
	{Cycle 2}; 
	\draw [thick,<->] (21.1,-2) -- (32,-2);
	\node[align=center,below] at (26.5,-2)%
	{Cycle 3};
	\draw [thick,<->,blue] (9.1,-1) -- (18,-1);
	\node[align=center,below] at (13.5,0)%
	{$U$};
	
	\node[align=center,left] at (6.5,2.5)%
	{\color{green} State $P$};
	\node[align=center,left] at (18.8,2)%
	{\color{red} State 0};
	\node[align=center,left] at (29.2,3.2)%
	{\color{green} State $P$};
	
	\draw [dashed,->] (8.3,5.6)  --(8.3,2.2);
	\draw [dashed,->] (20,5.6)  --(20.5,2.2);
	\node[align=center,left] at (21.5,6)%
	{\small Fluid stays at level $a$ for $\rm{Exp} (1)$ duration };

	\draw [dashed,->] (8.5,7.6) -- (5.2,5.2);
	\draw [dashed,->] (21,7.6) -- (27.6,5.1);
	\node[align=center,left] at (21.5,8)%
	{\small A successful arrival occurs above level $b$};
	\end{tikzpicture}
	\caption{Sample path of the random process $Z_f(t), t\geq 0$.\label{fig:samplePath3}}
\end{figure}
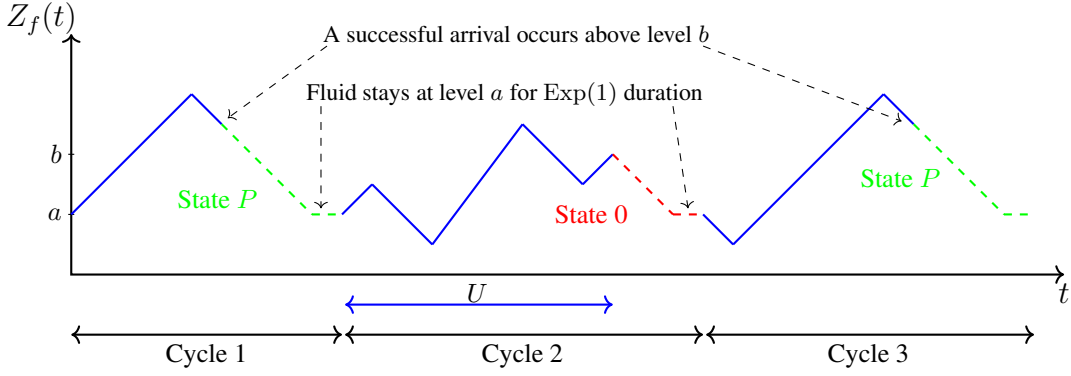

The following probability is of interest for the actual waiting time in the $MAP/M/s+G$ queue:
\begin{equation}
F_a^{a,b,\pi_0,\theta_0}(\tau)   = \Pr \{\tilde{\kappa}(a,b,\pi_0,\theta_0) < \tau \},
\label{realfirstpassagedistribution}
\end{equation}
which is used to express the likelihood whether any successful arrival within a time horizon $t$ will be subject to a waiting time above a certain threshold $b$. The deterministic time horizon, $\tau$, is again approximated by $U \sim PH(\beta,B)$ or $U \sim ME(\beta,B)$. For the purpose of obtaining the quantity in (\ref{realfirstpassagedistribution}), we introduce another auxiliary CFFQ $\mathbf{\tilde{Z}(t)}=(\tilde{Z}_f(t),\tilde{Z}_m(t)), t \geq 0$.
In this case, we let the modulating process $\tilde{Z}_m(t)$ to have the following state space:
\begin{equation} 
\tilde{Z}_{m}(t) \in \mathcal{\tilde{Z}} =  \{P\} \cup \{0\} \cup \left( \mathcal{S} \otimes 
\mathcal{M} \otimes \mathcal{U} \right).
\end{equation}
The size of the state space for $\tilde{Z}_{m}(t)$ is $2+\ell m (s+1)$ and similar to the previous section $\ell m (s+1)$ of these states correspond to the triple $(\sigma, i,j)$, $0\leq i \leq s$, $1 \leq j\leq m$ and $1 \leq \sigma \leq \ell$. In addition, there are two auxiliary reset states, namely state $P$ and state $0$. The fluid level process $\tilde{Z}_f(t)$ is a infinite capacity continuous-valued process in the interval $[0,\infty)]$ which is illustrated in Fig.~\ref{fig:samplePath3} for three consecutive cycles. Similar to the previous case, we intend ${\tilde{Z}_f(t)}$ to have a cyclic behavior. At each cycle, the fluid process starts from level $a$ and evolves according to the original fluid process but we also keep track of time. 
The reset states are used as follows: state $P$ is entered when a successful arrival occurs while the fluid process is above (or equal to) level $b$ and state $0$ is entered when the timer expires. In both of these states, fluid level is forced to level $a$ and stays at $a$ in the same state for an exponentially distributed duration with unit mean before a new cycle is initiated. With the infinitely many repetitions of cycles, these halt durations at level $a$ enable us to approximately obtain the quantity of interest given in (\ref{realfirstpassagedistribution}).


By enumerating the states of $\tilde{Z}_m(t)$ as 
\[P, 0, (0,1,1),(0,2,1),\ldots,(0,m,1),(1,0,1),\ldots,(1,m,1),\ldots,(s,m,1),(0,1,2),\ldots(s,m,\ell),
\]
the expressions for the matrix pair $\left( Q_{\tilde{Z}}(x),R_{\tilde{Z}}(x)\right) $ characterizing the CFFQ process $\mathbf{\tilde{Z}(t)}$ can now be given for the case $0\leq a<b$ (the case $0\leq b<a$ can be handled similarly). 
For this purpose, we first define the matrices $\tilde{Q}_Y(x)$ and $p(x)$:

\begin{align}
\tilde{Q}_Y(x) &=
\begin{cases}
Q_Y(x),& \text{if } 0\leq x<b,\\~\\
\left[ \begin{array}{c|c|c}
0_{m(s-1) \times m(s-1)} & 0_m & 0_m \\ \hline
0_{m \times m(s-1)} & C + Dg_a(x)  & 0_m\\ \hline
0_{m \times m(s-1)}& s \mu I_m & -s \mu I_m  
\end{array} \right],& \text{if } x\geq b,\\
\end{cases}
\label{tildeQYx} \\
\end{align}
and
\begin{equation}
p(x)= \begin{cases}
\left[ \begin{array}{c}
0_{m(s+1) \times 1}
\end{array} \right], & \text{if } 0\leq x<b,\\~\\
\left[ \begin{array}{c}
0_{m(s-1) \times 1}\\\hline
D g_s(x) 1_{m \times 1}\\ \hline
0_{m \times 1}
\end{array} \right], & \text{if } x\geq b.\\
\end{cases}
\label{p} 
\end{equation}
The level-dependent generator matrix $Q_{\tilde{Z}}(x)$ can now be written as:

\begin{eqnarray}
Q_{\tilde{Z}}(x) & = & \left[ \begin{array}{c|c|c}
-h(x) &0 &  h(x)(\alpha \otimes \pi_0 \otimes \theta_0) \\ \hline
0 &-h(x) & h(x)(\alpha \otimes \pi_0 \otimes \theta_0) \\ \hline
e_{\ell} \otimes p(x) &B^0 \otimes \tilde{e} & I_{\ell} \otimes \tilde{Q}_Y(x) + B \otimes\tilde{I}
\end{array} \right],
\label{tildeQ_YaMAP} \\
\end{eqnarray}
where
\begin{equation}
h(x)=0 \; {\rm for} \; x\neq a, h(a)=1, 
\end{equation}
and for the scenario $a> 0 $, the level-dependent drift matrix $R_{\tilde{Z}}(x)$ is written as:
\begin{equation}
R_{\tilde{Z}}(x) = 
\begin{cases} 
{\bf blockdiag}\{1,1,I_{\ell} \otimes R_Y(x)\}, & \text{if } x < a, \\
{\bf blockdiag}\{-1,-1,I_{\ell}\otimes R_Y(x)\}, & \text{if } x > a,   \\
{\bf blockdiag} \{0,0,I_{\ell}\otimes R_Y(a)\}, & \text{if } x = a. \\
\end{cases}
\label{tildeR_Z}
\end{equation}
For the specific scenario $a=0$, this drift matrix is slightly different.
\begin{equation}
R_{\tilde{Z}}(x) = 
\begin{cases} 
{\bf blockdiag}\{-1,-1,I_{\ell}\otimes R_Y(x)\}, & \text{if } x > 0, \\
{\bf blockdiag} \{0,0,I_{\ell}\otimes R_Y(0)\}, & \text{if } x = 0. \\
\end{cases}
\label{tildeR_Z2}
\end{equation}

The above expressions complete the matrix representation of the CFFQ $\mathbf{\tilde{Z}(t)}$ whose stationary solution can be found using the techniques briefed in Section~2.
Let $c_P(a)$ denote the steady-state probability mass at level $a$ for state $P$.
Also let $c_0(a)$ be the steady-state probability mass at level $a$ for state $0$. When a cycle terminates due to expiration of the timer, we end up at level $a$ during a visit to state $0$ for an exponentially distributed duration with unit parameter. For those cycles that terminate due to a successful arrival when the fluid level is above $b$ before the timer expires, we end up at level $a$ during a visit to state $P$ for an exponentially distributed duration with unit parameter. Based on this notation, we have 
\begin{equation}
F_a^{a,b,\pi_0,\theta_0}(\tau)  \approx  \frac{c_P(a)}{c_P(a)+c_0(a)}.
\label{realfirstpassage}
\end{equation}

\section{Numerical Examples}
\label{numerical}
In this section, analytical results are obtained with the proposed approach and are compared to the simulation results (with confidence intervals) for the purpose of validating the proposed analytical method. 
For the steady-state results, we opted to report the following performance metrics: $\Pr \{ W=0 \}, \Pr \{ W=0|\mathcal{S}\}, \Pr \{ \mathcal{A} \}, E(W|\mathcal{S}), \text{Var}(W|\mathcal{S}), F_{W|\mathcal{S},W>0}(.1),$ and $F_{W|\mathcal{S},W>0}(.2)$. In order to make the numerical experiments reproducible, we made our Matlab implementation available at \url{https://www.mathworks.com/matlabcentral/fileexchange/73599-map-m-s-g-solver-for-steady-state-and-first-passage-times} .

\subsection{Steady-state Results with Poisson Arrivals and Exponential Patience Times}
We set the packet arrival rate $\lambda=10$, number of servers $s=10$, service time $\mu=1$ and the abandonment time to be exponentially distributed with unit parameter for which case we also have the closed-form expressions given by Zelytn and Mandelbaum \cite{Zeltyn.2005} still requiring discretization and numerical integration.
Table~\ref{Tab:M/M/S+M} presents the performance metrics obtained for the $M/M/s+M$ scenario with the proposed analytical method with three different quantization levels $K=10,50,250$, the numerical results obtained by the method given in \cite{Zeltyn.2005}, and also the simulation results (with 95\% confidence intervals). We observe that the results obtained with the proposed analytical method are in line with those obtained with the existing exact method (up to four digits) when the number of quantization levels $K$ is sufficiently large. 

\begin{table}
	\centering
	\caption{Results for the $M/M/s+M$ scenario with $s=10$, $\lambda=10$, $\mu=1$, $E(A)=1$.}
		\begin{tabular}{lccccc}
			
			& \shortstack{Analysis \\ $K=10$} & \shortstack{Analysis \\ $K=50$} & \shortstack{Analysis  \\$K=250$ } & \shortstack{Ref.\\ \cite{Zeltyn.2005}}  & \shortstack{Simulation \\Results}                             \\ 
			\hline
			$\Pr \{ W=0 \}$      & 0.46000 & 0.45802 & 0.45794  & 0.45793               & $0.45829 \pm 0.00209$ \\
			$\Pr \{ W=0|\mathcal{S}\}$    & 0.52612 & 0.52353& 0.52343   & 0.52341             & $0.52375 \pm 0.00195$    \\
			$\Pr \{ \mathcal{A} \}$   &  0.12567 & 0.12513 & 0.12511    &   0.12511          & $0.12498 \pm 0.00084$     \\
			$E(W|\mathcal{S})$     & 0.11648 & 0.11500 & 0.11494       & 0.11494     &  $0.11491 \pm 0.00077$       \\
			$\text{Var}(W|\mathcal{S})$  & 0.03377 & 0.03310 & 0.03307   & 0.03307      &  $0.03299 \pm 0.00030$       \\
			$F_{W|\mathcal{S},W>0}(.1)$   & 0.28108 & 0.28100 & 0.28108      & 0.28107     &  $0.28060 \pm 0.00093$      \\
			$ F_{W|\mathcal{S},W>0}(.2)$   & 0.51155& 0.51285 & 0.51283      & 0.51283     & $0.51208 \pm 0.00145$    \\ \hline  
		\end{tabular}
		
		\label{Tab:M/M/S+M}
	
\end{table}

\subsection{Steady-state Results with MAP Arrivals and General Patience Times}
In this section, we consider the following particular patience time distributions some of which were also used in Jouini et al.  \cite{jouini_etal_IIE13}:
\begin{itemize}
	\item Exponential patience time with unit mean (denoted by $M$),
	\item Exponential patience time with balking (denoted by $M_B$) where the cdf of the patience time $A$ is given as 
	\begin{equation}
	F_A\left( x \right)=\frac{1}{3}\left( 1-{{e}^{-0.1x}} \right)+\frac{2}{3}, x \geq 0.
	\end{equation}
	\item Hyper-exponential patience time (denoted by $HE$) where
	\begin{equation}
	F_A\left( x \right)=\frac{1}{3}\left( 1-{{e}^{-0.1x}} \right)+\frac{2}{3}\left( 1-{{e}^{-x}} \right), x \geq 0
	\end{equation}
	\item Deterministic patience time (denoted by $D$) where the customers leave the system without getting served when they wait more than 0.5 time units,
	\item Weibull patience time (denoted by $W$) with rate 1 and shape parameter 3,
	\item Erlang-2 patience time (denoted by $E_2$) with mean 2,
	\item Erlang-3 patience time (denoted by $E_3$) with mean 3.
\end{itemize}
For the call arrival process, we use the arrival process proposed by Kawanishi and Takine \cite{Kawanishi.2016} which is a superposition
of $k, k \geq 1$ independent, homogeneous two-state Markov-modulated Poisson processes, each of which is characterized by an infinitesimal generator $Q_S$ of the
underlying two-state Markov chain and the arrival rate matrix $R_S$ given by
\[
Q_S= \begin{bmatrix}
-0.25 & 0.25 \\
1.0 & -1.0
\end{bmatrix},
R_S= \begin{bmatrix}
0.5 & 0\\
0 & 3.0
\end{bmatrix}.
\]
The overall arrival process is characterized with a matrix pair $(C,D)$ of size $k+1 \times k+1$ with the arrival rate $\lambda=k$. We refer to Kawanishi and Takine \cite{Kawanishi.2016} for more details for the matrix pair $(C,D)$. For the numerical examples of this section, we set $k=10$, $s=10$, and $\mu=1$. As before, three different values for the discretization parameter $K=10,50,250$ are used for discretizing the patience times. Obviously, discretization is not required for the deterministic abandonment case. The results of the proposed analytical approach are compared with the simulation results for the $MAP/M/s+M$, $MAP/M/s+M_B$, $MAP/M/s+HE$, $MAP/M/s+D$, $MAP/M/s+W$, $MAP/M/s+E_2$ and $MAP/M/s+E_3$ queueing systems in Tables~\ref{Tab:MAP/M/S+M}, \ref{Tab:MAP/M/S+MBalking}, \ref{Tab:MAP/M/S+HyperExpo}, \ref{Tab:MAP/M/S+D}, \ref{Tab:MAP/M/S+W}, \ref{Tab:MAP/M/S+E2}, and \ref{Tab:MAP/M/S+E3}, respectively. All the simulation results are presented along with the $\%95$ confidence intervals. Through these comparisons, it is shown that one can precisely and efficiently obtain the distribution of the steady-state queue waiting time as well as other performance metrics of interest using the proposed analytical method.


\begin{table}
	\centering
	\caption{Comparison of analytical results with simulations for the $MAP/M/s+M$ scenario.}
	\begin{tabular}{lcccc}
		
		& \shortstack{ Analysis \\ $K=10$} & \shortstack{Analysis \\ $K=50$} & \shortstack{Analysis \\ $K=250$}  & \shortstack{Simulation\\ Results}                             \\ 
		\hline
		$\Pr \{ W=0 \}$       & 0.43620& 0.43464 & 0.43458                 & $0.43473 \pm 0.00097$ \\
		$\Pr \{ W=0|\mathcal{S}\}$     & 0.51361 & 0.51152 & 0.51143              & $0.51158 \pm 0.00096$    \\
		$\Pr \{ \mathcal{A} \}$    &  0.15073 & 0.15029 & 0.15027          & $0.15023 \pm 0.00043$     \\
		$E(W|\mathcal{S})$     & 0.14046 & 0.13743 & 0.13737        &  $0.13718 \pm 0.00043$       \\
		$\text{Var}(W|\mathcal{S})$  & 0.12746 & 0.04456& 0.04451    &  $0.04437 \pm 0.00018$       \\
		$ F_{W|\mathcal{S},W>0}(.1)$     & 0.23845 & 0.23847 & 0.23854       &  $0.23854 \pm 0.00066$      \\
		$F_{W|\mathcal{S},W>0}(.2) $     & 0.44512 & 0.44644 & 0.44642      & $0.44661 \pm 0.00094$    \\ \hline   
	\end{tabular}
	\label{Tab:MAP/M/S+M}
\end{table}

\begin{table}
	\centering
	\caption{Comparison of analytical results with simulations for the $MAP/M/s+M_B$ scenario.}
	\begin{tabular}{lcccc}
		
		& \shortstack{Analysis\\ $ K=10$} & \shortstack{ Analysis \\ $K=50$} & \shortstack{Analysis\\ $K=250$}  & \shortstack{Simulation \\ Results}                             \\ 
		\hline
		$\Pr \{ W=0 \}$    & 0.69025 & 0.68969 &0.68968                & $0.68977 \pm 0.00034$ \\
		$\Pr \{ W=0|\mathcal{S}\}$    & 0.88516 &0.88428 & 0.88425              & $0.88430 \pm 0.00019$    \\
		$\Pr \{ \mathcal{A} \}$   &  0.22020 & 0.22006 & 0.22006          & $0.21999 \pm 0.00024$     \\
		$E(W|\mathcal{S})$       & 0.01563 & 0.01483 & 0.01480        &  $0.01481 \pm 0.00004$       \\
		$\text{Var}(W|\mathcal{S})$ & 0.00368 & 0.00344 & 0.00343    &  $0.00344 \pm 0.00001$       \\
		$F_{W|\mathcal{S},W>0}(.1) $    &0.53344&0.53371&0.53384       &  $0.53332\pm 0.00061$      \\
		$F_{W|\mathcal{S},W>0}(.2) $    & 0.78709 & 0.78846 & 0.78846      & $0.78832\pm 0.00055$ \\ \hline      
	\end{tabular}
	\label{Tab:MAP/M/S+MBalking}
\end{table}

\begin{table}
	\centering
	\caption{Comparison of analytical results with simulations for the $MAP/M/s+HE$ scenario.}
	\begin{tabular}{lcccc}
		
		& \shortstack{Analysis\\ $K=10$} & \shortstack{Analysis \\ $K=50$} & \shortstack{Analysis \\  $K=250$}  & \shortstack{Simulation \\Results   }                          \\ 
		\hline
		$\Pr \{ W=0 \}$       & 0.39314 & 0.39104 & 0.39096                 & $0.39136 \pm 0.00088$ \\
		$\Pr \{ W=0|\mathcal{S}\}$    & 0.45551 & 0.45276 & 0.45266             & $0.45311 \pm 0.00087$    \\
		$\Pr \{ \mathcal{A} \}$    &  0.13692 & 0.13632 & 0.13630          & $0.13629 \pm 0.00037$     \\
		$E(W|\mathcal{S})$     & 0.20596 & 0.20195& 0.20180        &  $0.20175 \pm 0.00061$       \\
		$\text{Var}(W|\mathcal{S})$  & 0.08667 & 0.08281 & 0.08266    &  $0.08266 \pm 0.00039$       \\
		$ F_{W|\mathcal{S},W>0}(.1)$     & 0.18329 & 0.18628 & 0.18639       &  $0.18610\pm 0.00054$      \\
		$F_{W|\mathcal{S},W>0}(.2)$     & 0.35371 & 0.35537 & 0.35547      & $0.35501\pm 0.00085$     \\ \hline  
	\end{tabular}
	\label{Tab:MAP/M/S+HyperExpo}
\end{table}

\begin{table}
	\centering
	\caption{Comparison of analytical results with simulations for the $MAP/M/s+D$ scenario.}
	\begin{tabular}{lcc}
		
		& Analysis  & Simulation Results                             \\ 
		\hline
		$\Pr \{ W=0\}$      & 0.37989         & $0.38040 \pm 0.00091$ \\
		$\Pr \{ W=0|\mathcal{S}\}$     & 0.43851      & $0.43893 \pm 0.00090$    \\
		$\Pr \{ \mathcal{A} \}$   &  0.13367               & $0.13334 \pm 0.00042$     \\
		$E(W|\mathcal{S})$      & 0.14990          &  $0.14974 \pm 0.00033$       \\
		$\text{Var}(W|\mathcal{S})$ & 0.02964          &  $0.02962 \pm 0.00004$       \\
		$ F_{W|\mathcal{S},W>0}(.1)$     & 0.17763   &  $0.17778 \pm 0.00047$      \\
		$F_{W|\mathcal{S},W>0}(.2)$    & 0.35825  & $0.35855 \pm 0.00072$    \\ \hline   
	\end{tabular}
	\label{Tab:MAP/M/S+D}
\end{table}

\begin{table}
	\centering
	\caption{Comparison of analytical results with simulations for the $MAP/M/s+W$ scenario.}
	\begin{tabular}{lcccc}
		
		& \shortstack{Analysis \\ $K=10$} & \shortstack{Analysis \\ $K=50$} & \shortstack{Analysis \\ $K=250$}  & \shortstack{Simulation \\ Results }                            \\ 
		\hline
		$\Pr \{ W=0 \}$      & 0.34776 & 0.33324 & 0.33138                & $0.33166 \pm 0.00097$ \\
		$\Pr \{ W=0|\mathcal{S}\}$     & 0.39620 & 0.37774 & 0.37538             & $0.37557 \pm 0.00098$    \\
		$\Pr \{ \mathcal{A} \}$   & 0.12227 & 0.11778 & 0.11721          & $0.11693 \pm 0.00042$     \\
		$E(W|\mathcal{S})$      & 0.31926& 0.25150 & 0.25138       &  $0.25104 \pm 0.00065$       \\
		$\text{Var}(W|\mathcal{S})$  & 3.91308& 0.12642 &0.08071    &  $0.07974 \pm 0.00018$       \\
		$F_{W|\mathcal{S},W>0}(.1) $     & 0.13862 & 0.13371 & 0.13361       &  $0.13401 \pm 0.00046$      \\
		$F_{W|\mathcal{S},W>0}(.2) $     & 0.27149 & 0.26618 & 0.26670      & $0.26723 \pm 0.00070$   \\ \hline    
	\end{tabular}
	\label{Tab:MAP/M/S+W}
\end{table}

\begin{table}
	\centering
	\caption{Comparison of analytical results with simulations for the $MAP/M/s+E_2$ scenario.}
	\begin{tabular}{lcccc}
		
		& \shortstack{Analysis \\ $K=10$} & \shortstack{Analysis \\ $K=50$} & \shortstack{Analysis \\ $K=250$}  & \shortstack{Simulation \\Results}                             \\ 
		\hline
		$\Pr \{ W=0 \}$     & 0.30664 & 0.29490 & 0.29380                 & $0.29336 \pm 0.00115$ \\
		$\Pr \{ W=0|\mathcal{S}\}$   & 0.34379& 0.32928 & 0.32793              & $0.32737 \pm 0.00115$    \\
		$\Pr \{ \mathcal{A} \}$   &  0.10806 & 0.10442& 0.10408         & $0.10393 \pm 0.00048$     \\
		$E(W|\mathcal{S})$      & 0.37254 & 0.37450 & 0.37482        &  $0.37475 \pm 0.00138$       \\
		$\text{Var}(W|\mathcal{S})$  & 0.17112 & 0.17499 & 0.17560    &  $0.17527 \pm 0.00075$       \\
		$F_{W|\mathcal{S},W>0}(.1) $    & 0.109761 & 0.10744 & 0.10785      &  $0.10782 \pm 0.00047$      \\
		$F_{W|\mathcal{S},W>0}(.2) $     & 0.21430 & 0.21334 & 0.21375     & $0.21371 \pm 0.00082$ \\ \hline      
	\end{tabular}
	\label{Tab:MAP/M/S+E2}
\end{table}

\begin{table}
	\centering
	\caption{Comparison of analytical results with simulations for the $MAP/M/s+E_3$ scenario.}
	\begin{tabular}{lcccc}
		
		& \shortstack{Analysis \\ $K=10$} & \shortstack{Analysis \\ $K=50$} & \shortstack{Analysis \\ $K=250$}  & \shortstack{ Simulation \\ Results}                             \\ 
		\hline
		$\Pr \{ W=0 \}$      & 0.24365 & 0.22187 &0.21925                & $0.21946 \pm 0.00106$ \\
		$\Pr \{ W=0|\mathcal{S}\}$     & 0.26659 & 0.24087 & 0.23780              & $0.23791 \pm 0.00108$    \\
		$\Pr \{ \mathcal{A} \}$   &  0.08603 & 0.07886 & 0.07801         & $0.07757 \pm 0.00044$     \\
		$E(W|\mathcal{S})$      &  0.63951 &0.65123 & 0.65284       &  $0.65137 \pm 0.00222$       \\
		$\text{Var}(W|\mathcal{S})$ &0.34263 & 0.38303 &0.38981   &  $0.39046 \pm 0.00157$       \\
		$F_{W|\mathcal{S},W>0}(.1) $     & 0.07563 & 0.06845 &0.06783       &  $0.06795 \pm 0.00035$      \\
		$F_{W|\mathcal{S},W>0}(.2) $    &0.14729&0.13513&0.13433      & $0.13458 \pm 0.00061$ \\ \hline      
	\end{tabular}
	\label{Tab:MAP/M/S+E3}
\end{table}
\subsection{Steady-state Results for Varying System Load and Number of Servers}
\label{varyingload}

In the previous example, the system load, number of servers, service time, and arrival rate were fixed, i.e., $ \rho=1, s=10, \mu=1, \lambda=10$. This time by fixing $s=10,\mu=1$, we first study the accuracy of the proposed solver for varying system loads for a particular 2-state MAP constructed by using the method presented by Mitchell \cite{mitchell_orl01}
which introduces auto-correlation into an arbitrary PH-type arrival process without changing its marginal interarrival time distribution.
Therefore, given a PH-type
distribution characterized by the pair $(v,T)$, then the
MAP defined by
\[
C = T, \;D = -(1 - \psi)Tev - \psi T, 0 \leq \psi < 1
\]
has a lag-$k$ autocorrelation between the $i^{th}$ and $(i+k)^{th}$ 
interarrival times in the form $c \psi^k, k\geq 1$ for some constant $c$. A large value of the
correlation parameter $\psi$ implies a slow decay of the lag-$k$
autocorrelation function and therefore strong correlation between successive interarrival times. When $\psi=0$, we obtain the uncorrelated phase-type process. In our numerical example, we first construct a hyper-exponential interarrival time distribution with desired mean $\lambda=s \mu \rho$ for a given system load $\rho$ with the squared coefficient of variation $SCV=16$ using the method of balanced means given by Tijms \cite{tijms_book86} and we incorporate
auto-correlation with parameter $\psi =0.95$ into this process as described above. 
The following piece-wise constant patience time distribution is used in this numerical example.
\begin{align}
g_a(x) = 
\begin{cases} 
0.1\lfloor x \rfloor, & \text{if } x < 10, \\
1, & \text{if } x \geq 10,  \\
\end{cases}
\label{piecewise}
\end{align}
where $\lfloor \cdot \rfloor$ denotes the integer part of the argument. The arising $MAP/M/s+G$ queue can easily be shown to reduce to the steady-state solution of an MRMFQ with $K=11$ regimes and with a state space size of 22. The performance metrics of interest are obtained with the proposed analytical method and are presented in Table~\ref{load} for five different values of the system load parameter $\rho$ along with simulation results given with $\%95$ confidence intervals. Except for one single $\Pr \{ \mathcal{A} \}$ result when $\rho=1.25$, all the obtained results lie inside the confidence intervals validating the accuracy of the proposed analytical method. 

As a second example, we generate a two-state MAP with $\lambda=9, \mu=10/s, \rho=0.9,\psi =0.95$ for varying number of servers $s=4,16,64,256$ along with the same patience time distribution in (\ref{piecewise}). The performance metrics are obtained with the proposed analytical method and are presented in Table~\ref{server} for varying number of servers, $s$, along with simulation results given with $\%95$ confidence intervals. The effectiveness of the proposed analytical method is clearly demonstrated by the obtained results that all lie inside the confidence intervals except for the two results for the metric $\text{Var}(W|\mathcal{S})$ corresponding to $s=64,256$.

\begin{table}[htb]
	\fontsize{9}{6}\addtolength{\tabcolsep}{-1pt}{
		\centering
		\caption{The performance metrics of interest obtained for the $MAP/M/s+G$ queue for five values of the system load parameter $\rho$. (M) stands for the proposed analytical method whereas (S) 
			represents simulation results with $\%95$ confidence intervals. }
		
		\vspace*{0.2cm}

			\begin{tabular}{llccccccc}
				
				& $\rho$ & {$\Pr \{ W=0 \}$} & {$\Pr \{ W=0|\mathcal{S}\} $ }&{$\Pr \{ \mathcal{A} \}$} &{	$E(W|\mathcal{S})$} & {$\text{Var}(W|\mathcal{S})$} 
				& \shortstack[l]{$F_{W|\mathcal{S},W>0}(.1)$} &
				\shortstack[l]{$F_{W|\mathcal{S},W>0}(.2)$} 
				\\ \hline
				
				&$0.5$&0.28657&0.29510&0.02892&0.56660&0.39702&0.08325&0.16263\\ 
				
				&$0.75$ &0.05244&0.07040&0.25516&2.80646&1.81523&0.00624&0.01250\\
				
				\textbf{(M)} &$1$&0.04286&0.07162&0.40160&4.05898&3.40400&0.00479&0.00958\\
				
				&$1.25$	&0.03779&0.07441&0.49218&4.75435&4.80526&0.00443&0.00887\\
				
				&$1.5$	&0.03399&0.07629&0.55444&5.18792&5.93479&0.00429&0.00858\\ \hline 
				\renewcommand{\arraystretch}{3}
				
				&$0.5$& \shortstack{\\ 0.28663 \\ $\pm$0.00195} &\shortstack{\\0.29515 \\ $\pm$0.00195} &\shortstack{\\0.02894 \\ $\pm$0.00032} &\shortstack{\\0.56601 \\$\pm$0.00378} 
				&\shortstack{\\0.39747 \\$\pm$0.00400} &\shortstack{\\0.08350 \\$\pm$0.00071} &\shortstack{\\ 0.16260 \\$\pm$0.00123} \\ 
				
				&$0.75$ &\shortstack{\\ 0.05278 \\$\pm$0.00072} & 	\shortstack{\\0.07081 \\$\pm$0.00091} &\shortstack{\\0.25496 \\$\pm$0.00080}
				&\shortstack{\\2.80434 \\$\pm$0.00859} &\shortstack{\\1.80946 \\$\pm$0.00954} &\shortstack{\\0.00621 \\$\pm$0.00011} 
				&\shortstack{\\0.01244 \\ $\pm$0.00021} \\ 
				
				\textbf{(S)}&$1$
				&\shortstack{\\ 0.04283 \\$\pm$0.00055} &\shortstack{0.07143 \\ $\pm$0.00083} &\shortstack{\\ 0.40089 \\$\pm$0.00089}
				&\shortstack{\\ 4.05265 \\$\pm$0.01114} &\shortstack{3.39910 \\$\pm$0.01855} &\shortstack{\\ 0.00482 \\ $\pm$0.00008}
				&\shortstack{0.00963 \\ $\pm$0.00014} \\
				
				&$1.25$	&\shortstack{\\ 0.03794 \\$\pm$0.00041}
				&\shortstack{\\ 0.07455 \\$\pm$0.00070} &\shortstack{0.49134 \\$\pm$0.00080} &\shortstack{\\ 4.74759 \\$\pm$0.01063} &\shortstack{4.79891 \\$\pm$0.01944}
				&\shortstack{\\ 0.00446 \\$\pm$0.00006} 
				&\shortstack{\\ 0.00890 \\$\pm$0.00010} \\
				
				&$1.5$	&\shortstack{\\ 0.03380 \\$\pm$0.00036} &\shortstack{0.07585 \\$\pm$0.00068} &\shortstack{\\ 0.55471 \\$\pm$0.00085}
				&\shortstack{\\ 5.19303 \\$\pm$0.01189} &\shortstack{5.91386 \\$\pm$0.02311} &\shortstack{\\ 0.00425 \\ $\pm$0.00006} 
				&\shortstack{\\ 0.00852 \\$\pm$0.00010} \\ \hline
				
			\end{tabular}
			\label{load}
		}
	
\end{table}

\begin{table}[htb]
	\fontsize{9}{6}\addtolength{\tabcolsep}{-1pt}{
		\centering
		\caption{The performance metrics of interest obtained for the $MAP/M/s+G$ queue for four different values of $s$. (M) stands for the proposed analytical method whereas (S) 
			represents simulation results with $\%95$ confidence intervals.}
		
		\vspace*{0.2cm}

			\begin{tabular}{llccccccc}
				
				& $s$ & {$\Pr \{ W=0 \}$} & {$\Pr \{ W=0|\mathcal{S}\} $ }&{$\Pr \{ \mathcal{A} \}$} &{	$E(W|\mathcal{S})$} & {$\text{Var}(W|\mathcal{S})$} 
				
				& \shortstack[l]{$F_{W|\mathcal{S},W>0}(.1) $} &
				\shortstack[l]{$F_{W|\mathcal{S},W>0}(.2) $} 
				\\ \hline
				
				&$4$&0.03464&0.05382&0.35633&3.71820&2.58239&0.00511&0.01021\\

				\multirow{2}{*}{\textbf{(M)}}&$16$&0.05624&0.08630&0.34836&3.59057&2.95201&0.00511&0.01021\\
				
				&$64$	&0.12901&0.19014&0.32150&3.18253&3.91515&0.00511&0.01021\\
				
				&$256$	&0.30909&0.41490&0.25503&2.29934&4.85938&0.00511&0.01021\\ \hline 
				\renewcommand{\arraystretch}{3}

				&$4$& \shortstack{\\ 0.03450\\ $\pm$0.00029} &\shortstack{\\0.05361 \\$\pm$0.00043} &\shortstack{\\0.35672 \\$\pm$0.00048} &\shortstack{\\3.72260 \\$\pm$0.00564} 
				&\shortstack{\\2.57604 \\$\pm$0.00898} &\shortstack{\\0.00513\\$\pm$0.00005} &\shortstack{\\ 0.01023 \\$\pm$0.00010} \\

				\multirow{2}{*}{\textbf{(S)}}&$16$
				&\shortstack{\\ 0.05598 \\$\pm$0.00045} &\shortstack{0.08588 \\ $\pm$0.00062} &\shortstack{\\ 0.34860 \\$\pm$0.00057}
				&\shortstack{\\ 3.59489 \\$\pm$0.00685} &\shortstack{2.93859 \\$\pm$0.01000} &\shortstack{\\ 0.00509 \\ $\pm$0.00005}
				&\shortstack{ 0.01013 \\ $\pm$0.00009} \\
				
				&$64$	&\shortstack{\\ 0.12956 \\$\pm$0.00098}
				&\shortstack{\\ 0.19071 \\$\pm$ 0.00126} &\shortstack{0.32109 \\$\pm$0.00068} &\shortstack{\\ 3.17783 \\$\pm$0.00838} &\shortstack{3.90686 \\$\pm$0.01138}
				&\shortstack{\\ 0.00512 \\$\pm$0.00006} 
				&\shortstack{\\ 0.01028 \\$\pm$0.00011} \\
				
				&$256$	&\shortstack{\\ 0.30933 \\$\pm$0.00224}
				&\shortstack{\\ 0.41486 \\$\pm$ 0.00240} &\shortstack{0.25508 \\$\pm$0.00113} &\shortstack{\\ 2.30105 \\$\pm$0.01228} &\shortstack{4.84398 \\$\pm$0.00892}
				&\shortstack{\\ 0.00510 \\$\pm$0.00006} 
				&\shortstack{\\ 0.01028 \\$\pm$0.00011} \\ \hline

			\end{tabular}
			\label{server}
		
	}
\end{table}

\subsection{First Passage Time Results for the Virtual and Actual Waiting Times}
In this example, first passage time distributions for the $MAP/M/s+G$ queue are considered where $s=10$, $\mu=1$ and the abandonment time distribution is piece-wise constant and is as given in (\ref{piecewise}). The arrival process is a 2-state MAP constructed as in Subsection~\ref{varyingload} for $\rho=0.99, SCV=16, \psi=0.95$ which gives rise to the following 2-state $MAP(C,D)$ with:
\begin{align}
C &=  \begin{bmatrix}
-19.1994  &0\\
0&-0.6006
\end{bmatrix},  \quad D =  \begin{bmatrix}
19.1703  &  0.0291\\
0.0291  &  0.5715
\end{bmatrix}
\end{align}
In order to approximate the deterministic time horizon $\tau \in \{1,5,25 \}$, we make use of Erlang-$\ell$ and CME-$\ell$ distributions for three values of the order parameter $\ell \in \{ 25,51,101 \}$. As for the initial condition, we assume that the system is empty at the beginning , i.e., $a=0$ and 
$\pi_0=\{1, 0_{1 \times 10}\}$.

The cdf of the virtual waiting time evaluated at time $\tau$ , $F_v^{a,b,\pi_0,\theta_0}(\tau)$, is given in tables~\ref{tab:10} and \ref{tab:11}, for the MAP initial vector $\theta_0=\{1, 0\}$ and $\theta_0=\{0, 1\}$, respectively. 
The cdf of the actual waiting time evaluated at time  $\tau$, $F_a^{a,b,\pi_0,\theta_0}(t)$ is given in tables~\ref{tab:12} and \ref{tab:13}, for the MAP initial vector $\theta_0=\{1, 0\}$ and $\theta_0=\{0, 1\}$, respectively, again for the same set of values for the parameter $b$. 
For each table, simulation results are presented along with $\%99$ confidence intervals. The main observation is that the proposed method works very accurately for all the studied scenarios but the CME-$\ell$ approximation being more effective than Erlang-$\ell$ for fixed $\ell$. This situation is more emphasized for relatively smaller values of $\tau$. We further observe that, even with a relatively small choice of the parameter $\ell$, very accurate performance figures can be obtained with CME-$\ell$ approximating the deterministic time horizon $\tau$. We believe that this observation is crucial since the computational complexity of the proposed approach is cubic in the parameter $\ell$. Furthermore, all the obtained results lie inside the confidence intervals for the case when $\tau$ is approximated with CME-$101$, validating the accuracy of the proposed analytical method.

\begin{table}[htb]

		\centering
		\caption{The cdf of the virtual waiting time evaluated at time $\tau$, $F_v^{a,b,\pi_0,\theta_0}(\tau)$, obtained by the proposed method using Erlang-$\ell$ and CME-$\ell$ for three values of $\ell$, for the scenario $\theta_0=\{1, 0\}$.}
		\label{tab:10} 
	
			\begin{tabular}{ccccccccc}
				&\multicolumn{1}{l}{}&\multicolumn{1}{c}{}&\multicolumn{3}{c}{Erlang-$\ell$} &\multicolumn{3}{c}{CME-$\ell$}\\  \noalign{\smallskip}
				\multicolumn{1}{c}{$\tau$} &\multicolumn{1}{c}{$b$} &\multicolumn{1}{c}{Simulation}&\multicolumn{1}{l}{$\ell=25$}&\multicolumn{1}{l}{$\ell=51$} &\multicolumn{1}{c}{$\ell=101$}&\multicolumn{1}{c}{$\ell=25$}&\multicolumn{1}{c}{$\ell=51$}&\multicolumn{1}{c}{$\ell=101$} \\ \hline \noalign{\smallskip}
				&$1/4$&0.61906$\pm$0.00027&0.59251&0.60529&0.61200&0.61705&0.61883&0.61922\\
				
				&$1/2$ &0.39325$\pm$0.00050&0.38744&0.39032&0.39185&0.39309&0.39345&0.39354\\
				
				$1$&$1$&0.11791$\pm$0.00065&0.13245&0.12551&0.12189&0.11922&0.11822&0.11800\\
				
				&$2$	&0.00323$\pm$0.00044&0.00607&0.00456&0.00388&0.00345&0.00328&0.00324\\ 
				
				&$4$	&0.00000$\pm$0.00002&0.00000&0.00000&0.00000&0.00000&0.00000&0.00000\\ \hline
				
				&$1/4$ &0.97700$\pm$0.00022&0.97692&0.97698&0.97700&0.97690&0.97699&0.97700\\
				
				&$1/2$ &0.96964$\pm$0.00029&0.96910&0.96949&0.96960&0.96953&0.96965&0.96967\\
				
				$5$&$1$&0.95339$\pm$0.00031&0.94898&0.95163&0.95257&0.95294&0.95326&0.95331\\
				
				&$2$	&0.86128$\pm$0.00039&0.82707&0.84443&0.85283&0.85859&0.86077&0.86124\\
				
				&$4$	&0.14516$\pm$0.00050&0.16177&0.15408&0.14986&0.14667&0.14544&0.14517\\ \hline
				
				&$1/4$ &0.98695$\pm$0.00016&0.98684&0.98691&0.98694&0.98693&0.84307&0.98697\\
				
				&$1/2$&0.98274$\pm$0.00020&0.98261&0.98270&0.98274&0.98273&0.98278&0.98278\\
				
				$25$&$1$&0.97408$\pm$0.00022&0.97381&0.97394&0.97400&0.97399&0.97405&0.97406\\
				
				&$2$	&0.95078$\pm$0.00035&0.95051&0.95073& 0.95084&0.95084&0.95093&0.95094\\
				
				&$4$	&0.86036$\pm$0.00035&0.85903&0.85979&0.86006&0.86011&0.86027&0.86030\\ \hline
			\end{tabular}

\end{table}

\begin{table}[hbt]

		\centering
		\caption{The cdf of the virtual waiting time evaluated at time $\tau$, $F_v^{a,b,\pi_0,\theta_0}(\tau)$, obtained by the proposed method using Erlang-$\ell$ and CME-$\ell$ for three values of $\ell$, for the scenario $\theta_0=\{0, 1\}$.}
		\label{tab:11} 
		\smallskip
	
			\begin{tabular}{ccccccccc}
				&\multicolumn{1}{l}{}&\multicolumn{1}{c}{}&\multicolumn{3}{c}{Erlang-$\ell$} &\multicolumn{3}{c}{CME-$\ell$}\\\noalign{\smallskip}
				\multicolumn{1}{l}{$\tau$} &\multicolumn{1}{c}{$b$} &\multicolumn{1}{c}{Simulation}&\multicolumn{1}{c}{$\ell=25$}&\multicolumn{1}{c}{$\ell=51$} &\multicolumn{1}{c}{$\ell=101$}&\multicolumn{1}{c}{$\ell=25$}&\multicolumn{1}{c}{$\ell=51$}&\multicolumn{1}{c}{$\ell=101$} \\ \hline \noalign{\smallskip}
				&$1/4$&0.00663$\pm$0.00010&0.00703&0.00681& 0.00671&0.00664&0.00661&0.00660\\
				
				&$1/2$ &0.00355$\pm$0.00006&0.00400&0.00377&0.00365&0.00357&0.00354&0.00353\\
				
				$1$&$1$&0.00081$\pm$0.00003&0.00111&0.00096&0.00089&0.00084&0.00082&0.00082\\
				
				&$2$	&0.00002$\pm$0.00001&0.00004&0.00002&0.00002&0.00002&0.00002&0.00002\\
				
				&$4$	&0.00000$\pm$0.00000&0.00000&0.00000&0.00000&0.00000&0.00000&0.00000\\ \hline
				
				&$1/4$ &0.11115$\pm$0.00032&0.11067&0.11085&0.11094&0.11100&0.11102&0.11103\\
				
				&$1/2$ &0.10346$\pm$0.00031&0.10305&0.10323&0.10332&0.10338&0.10340&0.10341\\
				
				$5$&$1$&0.08827$\pm$0.00026&0.08794&0.08809&0.08816&0.08823&0.08824&0.08825\\
				
				&$2$	&0.05324$\pm$0.00020&0.05382&0.05349&0.05334&0.05326&0.05322&0.05321\\
				
				&$4$	&0.00365$\pm$0.00005&0.00538&0.00453&0.00412&0.00382&0.00372&0.00369\\ \hline
				
				&$1/4$ &0.49669$\pm$0.00059&0.49161&0.49415&0.49537&0.49620&0.49653&0.49660\\
				
				&$1/2$&0.48996$\pm$0.00058&0.48497&0.48751&0.48872&0.48956&0.48989&0.48996\\
				
				$25$&$1$&0.47661$\pm$0.00058&0.47159&0.47410&0.47531&0.47614&0.47647&0.47654\\
				
				&$2$	&0.44349$\pm$0.00059&0.43858&0.44104& 0.44223&0.44304&0.44336&0.44343\\
				
				&$4$	&0.33702$\pm$0.00049&0.33222&0.33441&0.33547&0.33622&0.33649&0.33654\\ \hline

			\end{tabular}
	
\end{table}

\begin{table}[hbt]

		\centering
		\caption{The cdf of the actual waiting time evaluated at time $\tau$, $F_a^{a,b,\pi_0,\theta_0}(\tau)$, obtained by the proposed method using Erlang-$\ell$ and CME-$\ell$ for three values of $\ell$, for the scenario $\theta_0=\{1, 0\}$.}
		\label{tab:12} 
		\smallskip

			\begin{tabular}{ccccccccc}
				&\multicolumn{1}{l}{}&\multicolumn{1}{c}{}&\multicolumn{3}{c}{Erlang-$\ell$} &\multicolumn{3}{c}{CME-$\ell$}\\\noalign{\smallskip}
				\multicolumn{1}{l}{$\tau$} &\multicolumn{1}{c}{$b$} &\multicolumn{1}{c}{Simulation}&\multicolumn{1}{c}{$\ell=25$}&\multicolumn{1}{c}{$\ell=51$} &\multicolumn{1}{c}{$\ell=101$}&\multicolumn{1}{c}{$\ell=25$}&\multicolumn{1}{c}{$\ell=51$}&\multicolumn{1}{c}{$\ell=101$} \\ \hline \noalign{\smallskip}
		&$1/4$&0.52092$\pm$0.00051&0.50227&0.51121&0.51598&0.51964&0.52089&0.52117\\
	
		&$1/2$ &0.31345$\pm$0.00049&0.31514&0.31438&0.31397&0.31375&0.31358&0.31355\\\
	
		$1$&$1$&0.08271$\pm$0.00028&0.09792&0.09062&0.08685&0.08408&0.08305&0.08282\\
	
		&$2$	&0.00184$\pm$0.00004&0.00382&0.00273&0.00226&0.00197&0.00186&0.00183\\
	
		&$4$	&0.00000$\pm$0.00000&0.00000&0.00000&0.00000&0.00000&0.00000&0.00000\\ \hline
	
		&$1/4$ &0.97418$\pm$0.00020&0.97393&0.97406&0.97409&0.97400&0.97409&0.97411\\
	
		&$1/2$ &0.96674$\pm$0.00023&0.96574&0.96635&0.96652&0.96649&0.96662&0.96664\\
	
		$5$&$1$&0.94941$\pm$0.00023&0.94345&0.94692&0.94819&0.94876&0.94915&0.94922\\
	
		&$2$	&0.84264$\pm$0.00052&0.80496&0.82380&0.83310&0.83960&0.84204&0.84256\\
	
		&$4$	&0.11830$\pm$0.00040&0.13624&0.12782&0.12323&0.11975&0.11844&0.11815\\\hline
	
		&$1/4$ &0.98533$\pm$0.00015&0.98517&0.98524&0.98528&0.98527&0.98530&0.98531\\
	
		&$1/2$&0.98108$\pm$0.00018&0.98089&0.98098&0.98103&0.98102&0.98106&0.98107\\
	
		$25$&$1$&0.97205$\pm$0.00017&0.97180&0.97193&0.97200&0.97199&0.97205&0.97206\\
	
		&$2$	&0.94819$\pm$0.00020&0.94767&0.94789& 0.94800&0.94801&0.94810&0.94811\\
	
		&$4$	&0.85268$\pm$0.00038&0.85115&0.85203&0.85234&0.85240&0.85257&0.85260\\\cline{2-9}

	\end{tabular}

\end{table}

\begin{table}[hbt]

		\centering
		\caption{The cdf of the actual waiting time evaluated at time $\tau$, $F_a^{a,b,\pi_0,\theta_0}(\tau)$, obtained by the proposed method using Erlang-$\ell$ and CME-$\ell$ for three values of $\ell$, for the scenario $\theta_0=\{0, 1\}$.}
		\label{tab:13} 
		\smallskip

			\begin{tabular}{ccccccccc}
				&\multicolumn{1}{l}{}&\multicolumn{1}{c}{}&\multicolumn{3}{c}{Erlang-$\ell$} &\multicolumn{3}{c}{CME-$\ell$}\\\noalign{\smallskip}
				\multicolumn{1}{l}{$\tau$} &\multicolumn{1}{c}{$b$} &\multicolumn{1}{c}{Simulation}&\multicolumn{1}{c}{$\ell=25$}&\multicolumn{1}{c}{$\ell=51$} &\multicolumn{1}{c}{$\ell=101$}&\multicolumn{1}{c}{$\ell=25$}&\multicolumn{1}{c}{$\ell=51$}&\multicolumn{1}{c}{$\ell=101$} \\ \hline \noalign{\smallskip}
		&$1/4$&0.00502$\pm$0.00005&0.00553&0.00529&0.00517&0.00509&0.00506&0.00505\\
	
		&$1/2$ &0.00257$\pm$0.00004&0.00307&0.00283&0.00271&0.00263&0.00260&0.00260\\
	
		$1$&$1$&0.00053$\pm$0.00003&0.00078&0.00066&0.00060&0.00056&0.00054&0.00054\\
	
		&$2$	&0.00001$\pm$0.00000&0.00002&0.00002&0.00001&0.00001&0.00001&0.00001\\
	
		&$4$	&0.00000$\pm$0.00000&0.00000&0.00000&0.00000&0.00000&0.00000&0.00000\\ \hline
		
		&$1/4$ &0.10802$\pm$0.00030&0.10761&0.10780&0.10789&0.10795&0.10797&0.10798\\
	
		&$1/2$ &0.10030$\pm$0.00029&0.10000&0.10018&0.10026&0.10033&0.10035&0.10035\\
	
		$5$&$1$&0.08481$\pm$0.00026&0.08463&0.08476&0.08483&0.08489&0.08491&0.08491\\
	
		&$2$	&0.04944$\pm$0.00022&0.05027&0.04985&0.04966&0.04954&0.04949&0.04948\\
	
		&$4$	&0.00283$\pm$0.00007&0.00438&0.00362&0.00324&0.00298&0.00289&0.00287\\ \hline
	
		&$1/4$ &0.49362$\pm$0.00051&0.48896&0.49149&0.49271&0.49355&0.49387&0.49394\\
	
		&$1/2$&0.48698$\pm$0.00051&0.48230&0.48483&0.48605&0.48688&0.48721&0.48727\\
	
		$25$&$1$&0.47332$\pm$0.00051&0.46860&0.47111&0.47232&0.47315&0.47347&0.47354\\
	
		&$2$	&0.43941$\pm$0.00051&0.43475&0.43720&0.43839&0.43920&0.43951&0.43958\\
	
		&$4$	&0.32884$\pm$0.00048&0.32438&0.32654&0.32759&0.32832&0.32859&0.32865\\ \hline

	\end{tabular}

\end{table}

\section{Conclusions}
In this paper, the $MAP/M/s+G$ call center queueing model with generally distributed patience times is studied. Using sample-path arguments, finding the steady-state distribution of the virtual waiting time in the $MAP/M/s+G$ queue is shown to reduce to finding the steady-state solution of a properly constructed CFFQ. The proposed method is exact when the patience time is discrete-distributed. In case of continuous/hybrid patience times, the CFFQ needs to be approximated by an MRMFQ for which numerically stable and efficient algorithms are available.
Our numerical results show that the performance metrics of interest can accurately be obtained with a suitable number of discretization levels or equivalently the number of regimes of the MRMFQ. Moreover, obtaining the first passage time distribution for the virtual and actual waiting time for the same queue is shown to reduce to finding the steady-state solution of a larger dimensionality CFFQ where the deterministic time horizon needs to be approximated by Erlang-$\ell$ or CME-$\ell$ distributions. Through numerical examples, we show that the CME-$\ell$ distribution is more effective than Erlang-$\ell$ for fixed order parameter $\ell$ and the proposed approximate transient method with CME-$\ell$ being used to approximate the deterministic time horizon, leads us to obtain very accurately the first passage times for the $MAP/M/s+G$ call center queueing model for a wide range of system parameters. We believe that the methods described here can be used in dimensioning and dynamic provisioning of call centers.
Future work will consist of incorporation of PH-type models for service times, first passage time distributions for quantities of interest other than the virtual and actual waiting time, and the transient distribution of the $MAP/M/s+G$ queue at a given time.


%
%
%
%

\end{document}